\documentclass[preprint]{aastex}

\def\lsim{\mathrel{\rlap{\lower4pt\hbox{\hskip1pt$\sim$}}
    \raise1pt\hbox{$<$}}}                
\def\gsim{\mathrel{\rlap{\lower4pt\hbox{\hskip1pt$\sim$}}
    \raise1pt\hbox{$>$}}}                

\newcommand{\lya}{Ly$\alpha$}

\newcommand{\degsq}{{\mathrm{deg^{2}}}}

\usepackage{lscape}
\usepackage{graphicx}

\usepackage{natbib, epsfig}

\shorttitle{QLF at $z\sim 4$}
\begin{document}


\title{The Faint End of the Quasar Luminosity Function at $z\sim 4$\altaffilmark{1} }

\author{Eilat Glikman\altaffilmark{2,3,4}, Milan Bogosavljevi\'{c}\altaffilmark{2}, S.~G. Djorgovski\altaffilmark{2}, Daniel Stern\altaffilmark{5,6}, Arjun Dey\altaffilmark{7}, Buell T. Jannuzi\altaffilmark{7}, Ashish Mahabal\altaffilmark{2}}

\altaffiltext{1}{The data presented herein were obtained at the W.M. Keck
Observatory, which is operated as a scientific partnership among the
California Institute of Technology, the University of California and the
National Aeronautics and Space Administration. The Observatory was made
possible by the generous financial support of the W.M. Keck Foundation.}

\altaffiltext{2}{Astronomy Department, California Institute of Technology,
Pasadena, CA 91125}

\altaffiltext{3}{ Department of Physics and Yale Center for Astronomy and Astrophysics, Yale University, P.O. Box 208121, New Haven, CT 06520-8121; email: eilat.glikman@yale.edu}

\altaffiltext{4}{NSF Astronomy and Astrophysics Postdoctoral Fellow}

\altaffiltext{5}{Jet Propulsion Laboratory, Mail Stop 169-506 Pasadena, CA 91109}

\altaffiltext{6}{Visiting Astronomer, Kitt Peak National Observatory, National Optical Astronomy Observatory, which is operated by the Association of Universities for Research in Astronomy (AURA) under cooperative agreement with the National Science Foundation. }

\altaffiltext{7}{National Optical Astronomy Observatory, 950 N. Cherry Ave., Tucson, AZ 85719}

\begin{abstract}
The evolution of the quasar luminosity function is one of the basic cosmological measures providing insight into structure formation and mass assembly in the Universe.  We have conducted a spectroscopic survey to find faint quasars $(-26.0 < M_{1450} < -22.0)$ at redshifts $z=3.8-5.2$ in order to measure the faint end of the quasar luminosity function at these early times.  Using available optical imaging data from portions of the NOAO Deep Wide-Field Survey and the Deep Lens Survey, we have color-selected quasar candidates in a total area of 3.76 deg$^2$.  Thirty candidates have $R\leq 23$ mags. We conducted spectroscopic followup for 28 of our candidates and found 23 QSOs, 21 of which are reported here for the first time, in the $3.74 < z <5.06$ redshift range.  We estimate our survey completeness through detailed Monte Carlo simulations and derive the first measurement of the density of quasars in this magnitude and redshift interval.  We find that the binned luminosity function is somewhat affected by the K-correction used to compute the rest-frame absolute magnitude at 1450\AA. Considering only our $R\leq 23$ sample, the best-fit single power-law ($\Phi \propto L^\beta$) gives a faint-end slope $\beta = -1.6\pm0.2$.  If we consider our larger, but highly incomplete sample going one magnitude fainter, we measure a steeper faint-end slope $-2 < \beta < -2.5$.  In all cases, we consistently find faint-end slopes that are steeper than expected based on measurements at $z\sim 3$.  We combine our sample with bright quasars from the Sloan Digital Sky Survey to derive parameters for a double-power-law luminosity function.  Our best fit finds a bright-end slope, $\alpha = -2.4\pm0.2$,  and faint-end slope, $\beta = -2.3\pm0.2$, without a well-constrained break luminosity. This is effectively a single power-law, with $\beta = -2.7\pm0.1$.  We use these results to place limits on the amount of ultraviolet radiation produced by quasars and find that quasars are able to ionize the intergalactic medium at these redshifts.
\end{abstract}

\section{Introduction}

Understanding the evolution of quasars has been a subject of great importance since their discovery over four decades ago.  In particular, there is now substantial, multifaceted, and growing evidence for a correlation between the formation and evolution of galaxies and their central supermassive black holes (\citealt{Magorrian98,Ferrarese00}; also, see the proceedings edited by \citealt{Ho04}, or the review by \citealt{Djorgovski05}, and references therein).  Studies of the evolving QSO populations are important not only for their own sake, but also for providing insights into the formation and evolution of massive galaxies in general.

While quasars represent a relatively minor contributor ($\sim10\%$) to the overall energetics of the post-recombination universe, they dominate at high energies, and their radiative and mechanical feedback may significantly affect the formation and growth of their hosts and companions \citep{Silk98,Wyithe03b}.   Quasars are a significant contributor to the metagalactic ionizing radiation field at any redshift, although their role relative to that of star-forming galaxies has likely changed over the history of the Universe.  Quasars might still be important producers of the metagalctic UV radiation at high redshifts ($z \gtrsim 4-5$) as one approaches the end of the reionization era.  

Although we observe a decline in the density of quasars by a factor of $\sim 40$ between the peak of quasar activity at $z\sim 2.5$ and the end of reionization at $z\sim 6$ \citep{Fan01b}, this evolution has been traced only by the most luminous sources at redshifts beyond $z\sim 3$.   The primary observable which constrains the evolution of quasar populations and their effects on their environment is the quasar luminosity function (QLF) as a function of redshift.  At low redshifts, the QLF is well-represented by a double  power-law, $\Phi(L) = \Phi_* / L_* [(L/L_*)^\alpha + (L/L_*)^\beta]^{-1}$\citep{Boyle88,Pei95}.  The bright-end slope, $\alpha$, has typical measured values around $-3.4$ at redshifts $z<2.5$, and flattens to $\alpha \approx -2.5$ at $z\sim 4$ \citep{Fan01a}. The faint-end slope, $\beta$, is typically measured to be around $-1.7$ at low redshifts ($z \lesssim1$), although \citet{Hunt04} find a shallower value of $\beta \approx -1.2$ at $z\sim 3$.  

There has been considerable progress recently in the theoretical understanding of the shape of the QLF and its evolution.  The emerging picture is a complex interplay of QSO lifetimes and luminosity histories, powered by dissipative mergers, with the bright end of the QLF dominated by short-lived phases of QSOs radiating near the peak of their luminosities and the faint end determined by the distribution of QSO fueling lifetimes and feedback \citep{Hopkins05a,Hopkins06b,Hopkins06c}.  Thus, in this model the faint end of the QLF at high redshifts represents a critical observational constraint on the early formation history of massive black holes, their contribution to the reionization, and feedback processes affecting the formation of their host galaxies.

The true shape of the QLF at $z>4$ is currently unknown, largely due to the flux limits of most large-area surveys to date.  The recent availability of deep, relatively wide-field, multicolor, optical surveys in the public domain have enabled the search for faint quasars at these high redshifts and the determination of the faint end of their QLF.  In this paper, we utilize the Deep Lens Survey \citep[DLS;][]{Wittman02} and the NOAO Deep Wide-Field Survey \citep[NDWFS;][]{Jannuzi99} which go $\sim 4$ magnitudes deeper than the Sloan Digital Sky Survey (SDSS) and other large area sky surveys, to measure the faint end of the QLF at $z\sim 4$.  

The organization of the paper is as follows: in Section 2, we review the NDWFS and DLS survey data that were used for photometric candidate selection. Section 3 presents the simulations of quasar colors we have performed to obtain the QSO candidate color selection criteria. In Section 4, we describe the followup spectroscopy observations. In Section 5 we compute the QLF and analyze its shape in conjunction with published measurements at brighter magnitudes.  In Section 6 we examine the implications of our QLF by computing the contribution of quasars to the UV radiation field at $z\sim 4$.  We discuss our results in Section 7.  We use standard cosmological parameters throughout the paper: $H_0=70$ km s$^{-1}$ Mpc$^{-1}$, $\Omega_M=0.30$, and $\Omega_\Lambda=0.70$.

\section{Candidate Selection \label{sec:selection}}

We select our quasar candidates using the standard technique of looking for objects whose colors are outliers from the stellar locus.  At $z\sim 4$, our redshift of interest, $B$, $R$, and $I$ (or $z$) effectively and efficiently separate quasars from stars \citep{Kennefick95a,Kennefick95b,Richards02}.  We make use of publicly available deep imaging data from the NDWFS and the DLS, which we describe below.

\subsection{NOAO Deep Wide-Field Survey}

NDWFS is a deep imaging survey of two 9.3 deg$^2$ fields, Bo\"{o}tes and Cetus, in three optical and infrared bands ($B_W$, $R$, $I$ and $J$, $H$, $K$).  For our candidate selection, we obtained imaging data in the Bo\"{o}tes field from the Third Data Release (DR3) which is publicly available from the NOAO Science Archive\footnote{\url{http://www.archive.noao.edu/ndwfs/}}.  

\begin{figure}
   \plotone{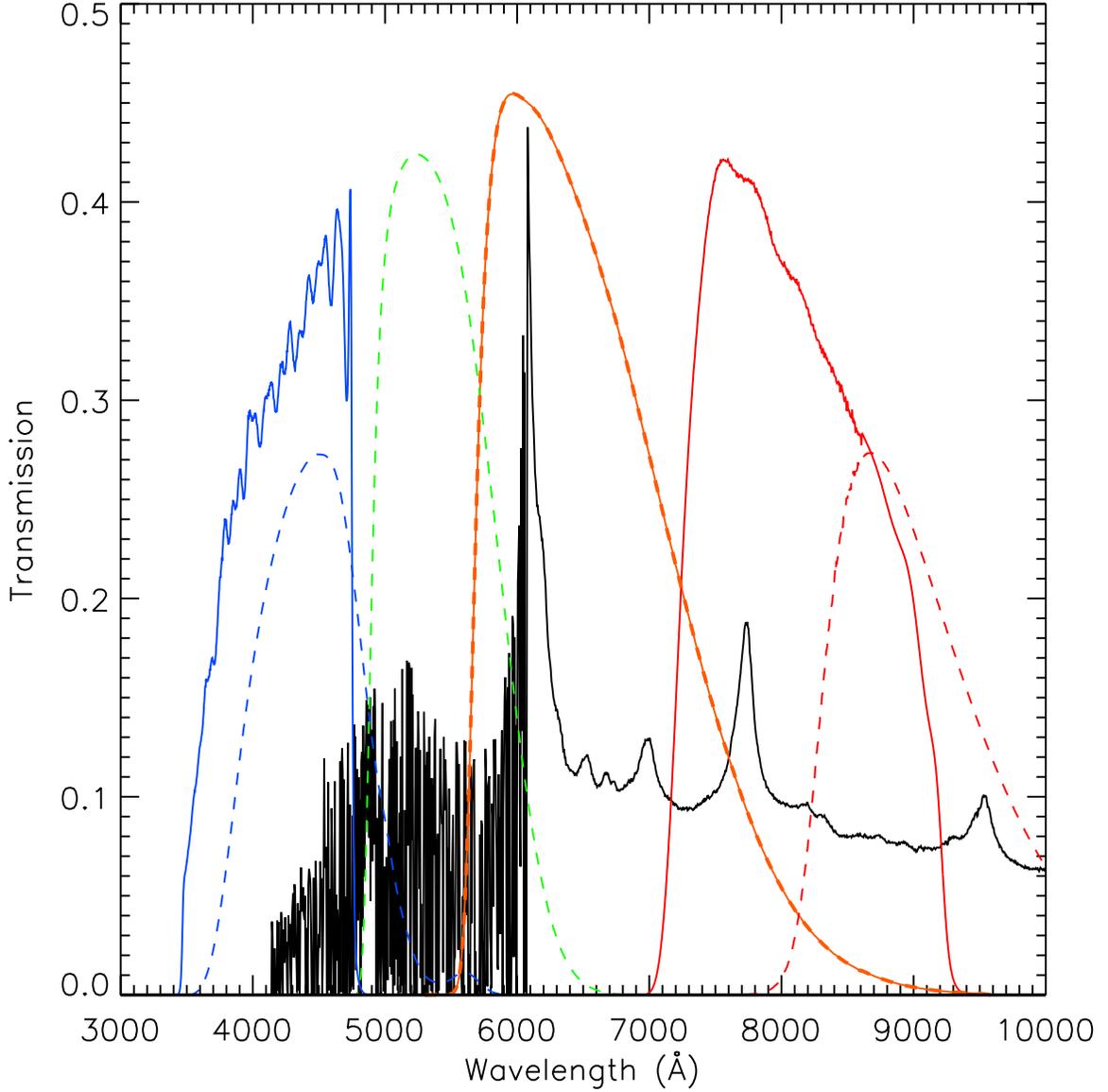}
   \caption
   {Effective throughput curves for the filters used in our quasar color-selection, which include the telescope and camera throughputs but do not include atmospheric absorption.  The NDWFS $B_W$,$R$, and $I$ filter curves are shown in solid blue, orange and red solid lines, respectively.  The DLS $B$, $V$, $R$ (which is identical to the NDWFS $R$ filter), and $z$ filters are shown in blue, green, orange and red dashed lines, respectively.  We overplot a simulated QSO at z=4.0.}
   \label{fig:nfilters}
 \end{figure}

The custom $B_W$ filter has high transmission over most of the wavelength range commonly covered by $U$ and $B$ filters, as can be seen in Figure ~\ref{fig:nfilters}. Optical imaging for the data available in NDWFS DR3 was obtained with the  4 meter Kitt Peak National Observatory (KPNO) Mayall telescope and MOSAIC-I Wide Field Imager. The Bo\"{o}tes field is split into 27 partially overlapping subfields, each roughly 36\arcmin$\times$36\arcmin\ on the sky, with imaging products delivered on a common (tangent-projected) scale of 0\farcs258 pixel$^{-1}$. The survey design aimed for 5$\sigma$ point source detection limits of $B_W\approx 26.6$, $R \approx 26.0$ and $I \approx 26.0$ (AB mags).

We use optical ($B_W$, $R$ and $I$) imaging for five of the 27 NDWFS subfields, selected to optimize the seeing conditions and total exposure time in all three filters.  The five subfields cover a total area of 1.71 $\degsq$ and their details are summarized in Table ~\ref{tabl:ndwfs_fields}.

The $B_W$, $R$, and $I$ images for a given subfield have different sizes.  We registered the images and trimmed them to their common overlapping area (listed in column 4 of Table \ref{tabl:ndwfs_fields}).  We then extracted source catalogs using SExtractor version 2.4.4 \citep{Bertin96} in dual image mode with $R$-band as the detection image and forcing the measurements in $B_W$- and $I$-bands. For detection in SExtractor, the $R$-band image was smoothed with a 3 pixel wide Gaussian filter, and the detection threshold was set to 2$\sigma$ of the sky noise. 

Photometry was performed in 3\arcsec\ diameter apertures. To compensate for differences in seeing between the three bands (see Table \ref{tabl:ndwfs_fields}), this aperture magnitude was corrected to ``total'' magnitude for each band separately by subtracting an aperture correction.  To determine the aperture correction in each image, we compared the isophotal magnitude for $\gsim$200 bright stellar objects to their aperture magnitude and measured the median difference.  This correction is typically $\lsim$ 0.1 mag in every band.  We also correct all the magnitudes for interstellar extinction using the dust map from \citet{SFD98}. For our selection, we work with these extinction and aperture-corrected magnitudes in the $AB_{\nu}$ system, converting the published Vega-based NDWFS zeropoints to $AB_{\nu}$ magnitudes by adding 0.0, 0.26 and 0.53 mag to the $B_W$, $R$ and $I$ zeropoints, respectively. 

\subsection{Deep Lens Survey}

DLS is closely related to the NDWFS in that the optical imaging has been obtained on the same 4 meter Mayall Telescope (at KPNO) and Blanco Telescope (at CTIO) using the same wide-field optical imagers Mosaic-I and Mosaic-II as the NDWFS. DLS imaging will ultimately cover a total of 28 deg$^2$ in four optical bands ($B$, $V$, $R$ and $z'$), separated on the sky into seven fields of 4 deg$^2$ (labeled F1 - F7). Each of these fields in turn is divided into 9 subfields, or ``pointings", of $35.1\arcmin \times 35.1 \arcmin$.  The final image products released by the DLS are already registered in pixel space.  Since the scale for these data is also 0\farcs258 pixel$^{-1}$, the final images have sizes $8192\times 8192$ pixels.  The DLS observational strategy was to obtain the $R$-band images on the best of the observing nights, while the $B$, $V$ and $z'$ data were taken in conditions of mixed relative quality.  In addition, the survey was designed to reach 12,000 seconds ($20\times 600$ seconds) in the $B$, $V$, and $z'$ filters and 18,000 seconds ($20\times 900$ seconds) in the $R$-band.  

We used the images from the third public data release of DLS data for 6 pointings, covering a total area of 2.05 $\degsq$.  The imaging data of the NDWFS and DLS differ in principle only in the filters used and fields surveyed; therefore, we applied the exact same procedure and parameters for catalog extraction as for the NDWFS data, including thresholds, apertures, and subsequent seeing and extinction corrections (see \S 2.1). The positions and several properties of the field pointings used in our survey are presented in Table \ref{tab:dls_fields}.

We extracted catalogs using SExtractor in dual image mode, with the $R$-band images as the detection image.  Photometric measurements were forced on the same pixel positions in the $B$, $V$ and $z'$ images.  The DLS data provide photometric zero-points.  The $B$, $V$, and $R$ images are calibrated to the Vega system, while the $z'$-band images are calibrated to the SDSS $AB_{\nu}$ system.  We convert from Vega magnitudes to $AB_{\nu}$ by adding $-0.09$, 0.0, and 0.2 to the $B$, $V$, and $R$ zero-points, respectively.  We also correct all the magnitudes for interstellar extinction using the dust map from \citet{SFD98}.  Although our main selection filters are $B$, $R$ and $z'$, we generated catalogs for the $V$-band data to further discriminate between our candidates and increase our efficiency, as we describe below.  

\subsection{ Field-to-field Consistency}

We examined the position of the stellar locus in the $(R-z')\ vs.\ (B-R)$ and $(R-I)\ vs.\ (B_W-R)$ color-color space for the DLS and NDWFS data, respectively, to verify the consistency of the zero-points. We found discrepancies between the fields that, while small, are larger than the expected photometric errors.  

\begin{figure}
      \plottwo{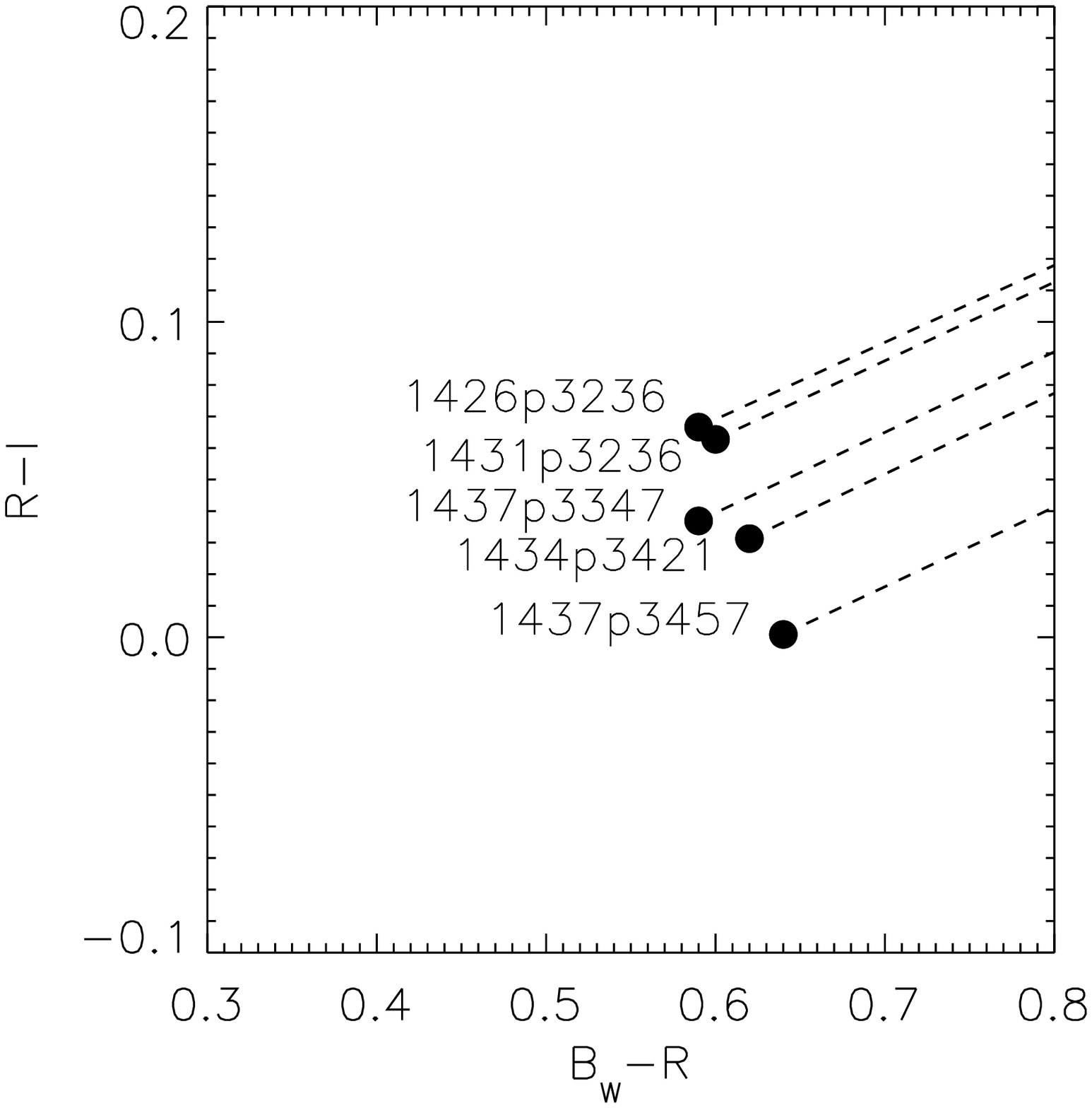}{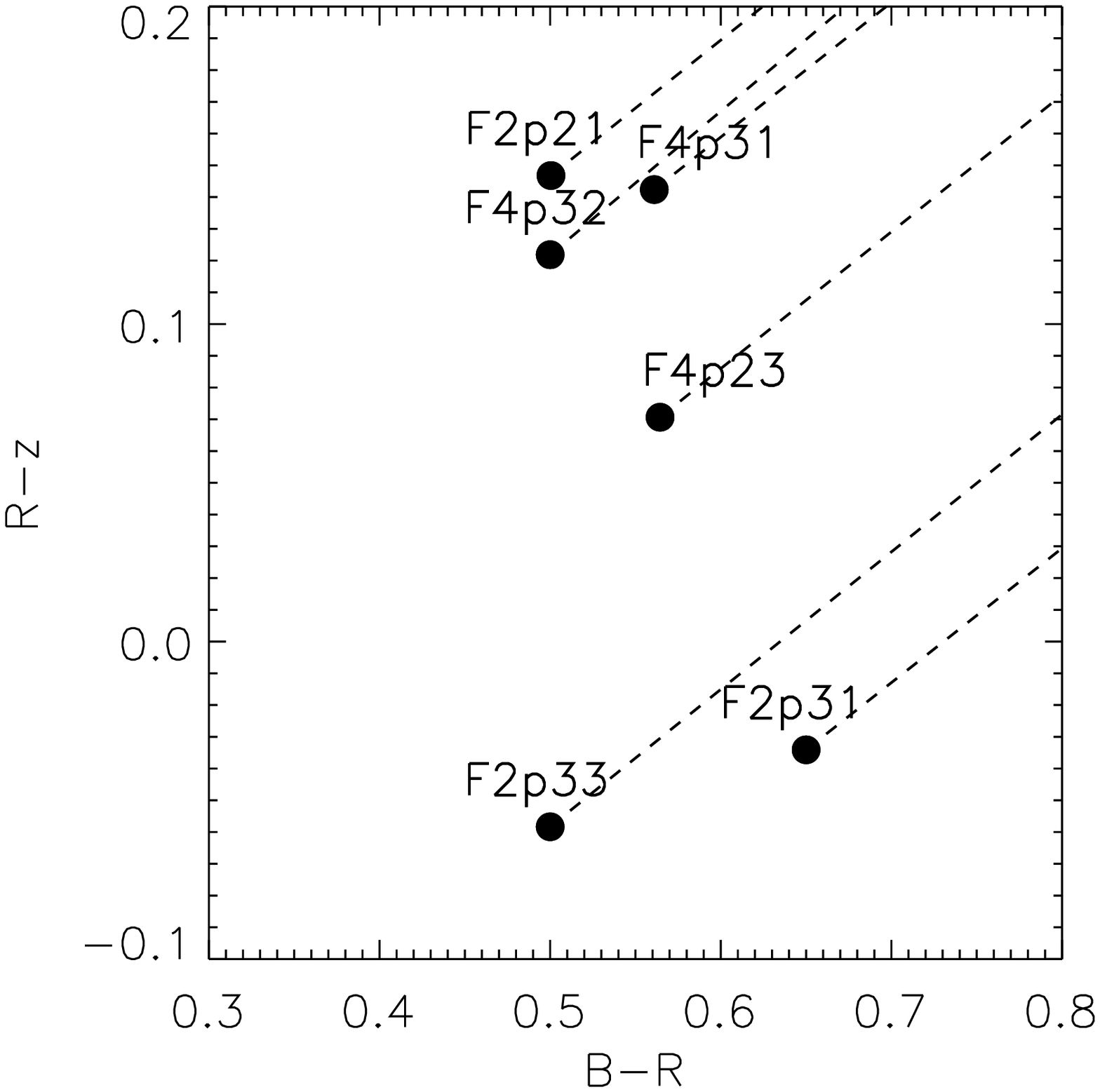}
\caption{Positions of the stellar locus color-tracks for the five NDWFS fields (left) and the six DLS fields (right). The filled circles indicate the origin of the stellar track and the dashed line traces the slope of the locus.  Each track is labeled by its corresponding field.}
   \label{fig:ccpos}
\end{figure}

To determine the zero-point shift required to bring our photometry onto a standard system we invoke a method used in the photometric quality assessment of SDSS.  Following \citet{Ivezic04}, we defined a principle axis in the aforementioned color-diagrams that follows the main sequence by iteratively fitting a line through the observed locus of well-detected stellar objects in each field.  We determined the origin of the stellar locus from the drop in density of stars along the determined track.  Figure \ref{fig:ccpos} shows the the positions of the main sequence color tracks for the five NDWFS fields on the left and the six DLS fields in our survey on the right.  The dashed lines indicate the principal axes, tracking that stellar locus, and the circles represent the origins of the locus.  

Two DLS fields, F2p31 and F2p33, have bluer $(R-z')$ colors, and F2p31 is redder than the other fields in $(B-R)$.  In addition, the photometric quality of F2p31 may be slightly poorer that the other five DLS fields as the stellar locus has an rms of $\sim 6\%$ compared with $3\%$ for the remaining fields.  We apply a correction to the $z'$-band zero-points of $-0.15$ and $-0.18$ to F2p31 and F2p33, respectively.  We also apply a $z'$-band zero-point shift of $-0.05$ magnitudes to F2p31.  The NDWFS data were consistent within photometric errors; therefore no correction was applied to those fields.

\section {Color Selection Criteria Using Simulated Quasar Spectra \label{sec:specsim}}

To define color selection criteria for quasar candidates at $z \gsim 4$ and determine our selection function, we simulated quasar spectra following the general procedure outlined in ~\citet{Fan99} and \citet{Richards06} and calculated their colors for the NDWFS and DLS filter combinations.  We generated a library of simulated quasar spectra, starting with a quasar template spectrum constructed from a combination of the {\it Hubble Space Telescope} ({\em HST}) radio-quiet quasar composite spectrum  from \citet{Telfer02} which covers $300-2461$\AA\ in the quasar rest frame and the SDSS quasar composite spectrum ~\citep{VandenBerk01} spanning rest-frame $800-9200$\AA.  This combined template uses the {\em HST} composite below 1275{\AA} and the SDSS composite above 2000{\AA}.  The spectra are averaged in the overlap region of $1275-2000$\AA\footnote{This combination is available at http://www.pha.jhu.edu/\~{}rt19/composite/.}.

We simulated the attenuation of each model spectrum by intervening neutral hydrogen absorbers by shifting it to a given redshift and creating random realizations of a population of discrete absorbers following the procedure outlined in ~\citet{Bershady99}.  The absorbers attenuate the template quasar spectrum with their Lyman series absorption lines and the Lyman limit break. Two separate distributions of absorbers are used, divided up according to their column density (N$_{\rm HI}$), corresponding to ``weak" absorbers and Lyman Limit Systems. We did not attempt to simulate the effect of any dust reddening or absorption by neutral hydrogen intrinsic to the quasar host galaxy.

We account for the intrinsic diversity of quasar spectral properties by varying the power-law slope of the continuum and the strength of the {\lya} emission line.  At UV-optical rest-frame wavelengths, the QSO continuum is best described by a broken power-law.  Shortward of {\lya}, we assign the spectrum a power-law index drawn from a Gaussian distribution centered on $\alpha_1 = -1.57$ with a standard deviation, $\sigma_1$, of $0.17$ \citep[$f_\nu \propto \nu^\alpha$;][]{Telfer02}.  Longward of {\lya} we assign the continuum a power-law index drawn from another Gaussian distribution centered on $\alpha_2=-0.5$ with $\sigma_2=0.30$ \citep{VandenBerk01,Richards03}.  We assume that the values of the two spectral slopes are uncorrelated. In addition, we vary the equivalent width, EW, of the {\lya} emission line assuming a Gaussian distribution of EW=$65\pm34${\AA} ~\citep{Wilkes86}, and again assume that it is uncorrelated with any of the other spectral properties. 

\begin{figure}
\plottwo{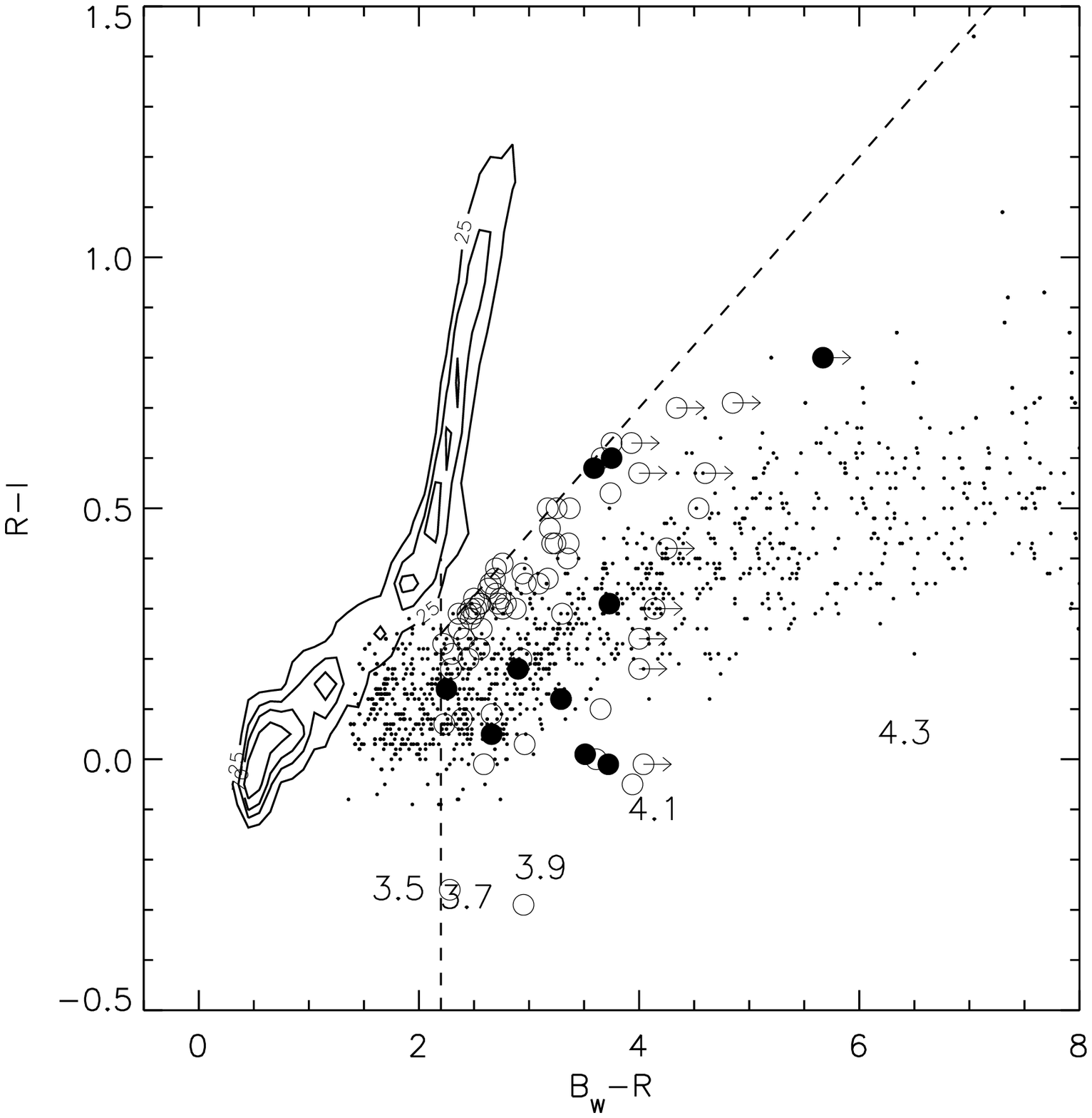}{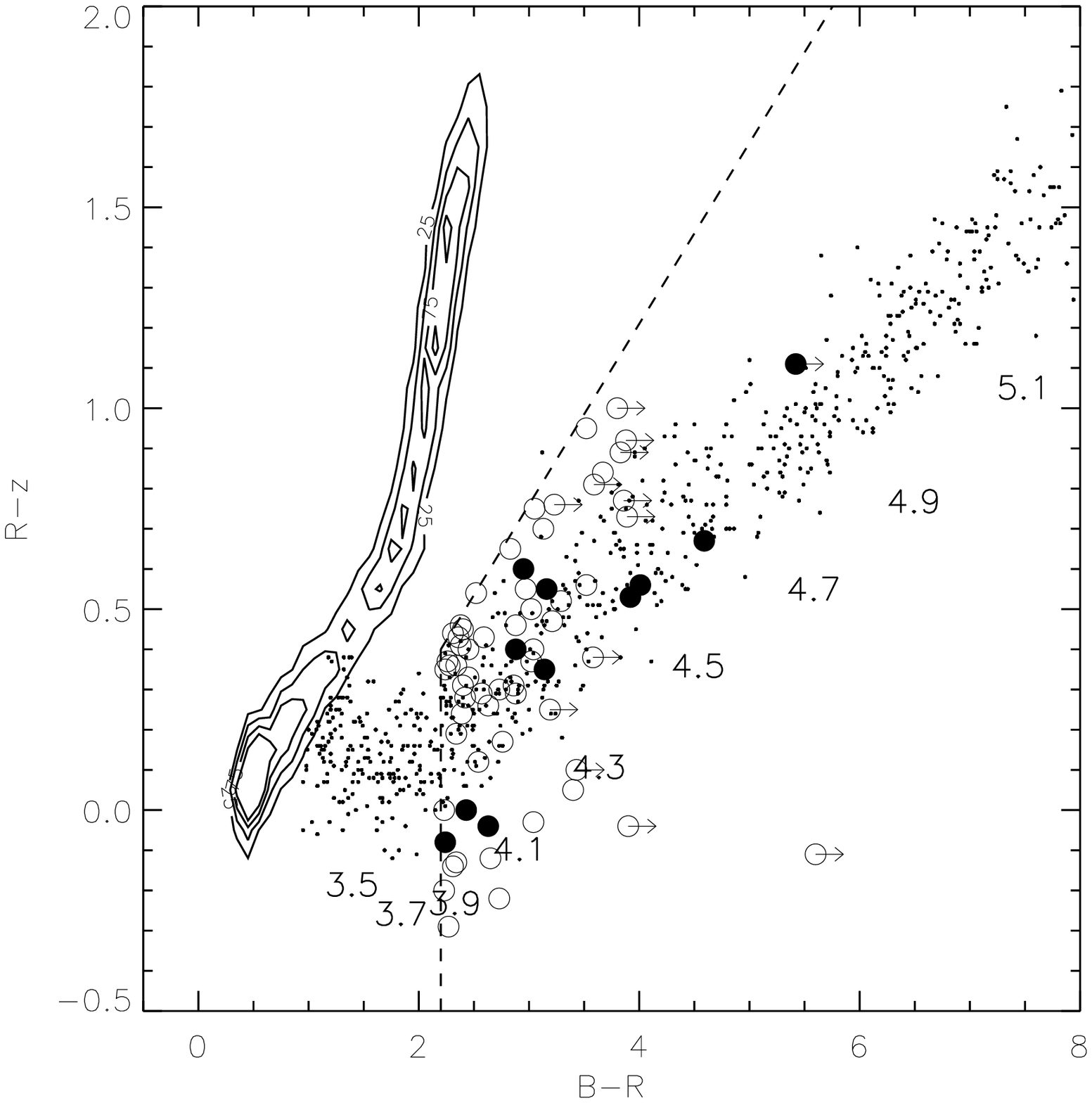}
\caption{Color-color diagram used in our quasar selection.  Small dots are the model quasar colors in the NDWFS $B_W$, $R$, $I$ filters (left) and the DLS $B$, $R$, $z'$ filters (right).  Density contours show the stellar locus using all stellar objects in each field.  The dashed lines show our color criteria for selecting quasars in each sample. The open circles show the candidates which passed our candidate selection criteria and the filled circles are confirmed quasars.}
\label{fig:colortracks}
\end{figure}

At each $\Delta z=0.1$ in the redshift range $3.5 < z < 5.2$, we created 200 simulated quasar spectra, each with a single drawing from distributions of the two power-law continuum slopes, the \lya\ emission line EW, and one realization of neutral hydrogen intervening absorption systems. Convolving these spectra with the filter responses for the NDWFS and DLS filters ($B_W$, $R$, $I$ and $B$, $V$, $R$, $z'$, where the two $R$ filters are identical) we obtain simulated quasar ``color tracks'' shown by the dots in Figure \ref{fig:colortracks}. 

Based on the colors of our model quasar spectra we adopt a set of color cuts for each of our datasets, NDWFS and DLS.  These are shown with dashed lines in Figure \ref{fig:colortracks}.  For the NDWFS survey, we impose the following selection criteria:
\begin{equation}
(B_W - R) \geq 2.2 
\end{equation}
and
\begin{equation}
(R - I) \leq 0.25 \times (B_W - R) - 0.3. 
\end{equation}
For the DLS survey, we use the following color cuts:
\begin{equation}
(B - R) \geq 2.2
\end{equation}
and
\begin{equation}
(R - z) \leq 0.45 \times (B - R) - 0.59.
\end{equation}
The DLS survey provides a fourth filter, $V$, which we use to impose an additional color criterion, based on the colors of the aforementioned simulations:
\begin{equation}
(B - V) \geq 1.0.
\end{equation}

\subsection{Morphological Criteria} 

\begin{figure}
\plottwo{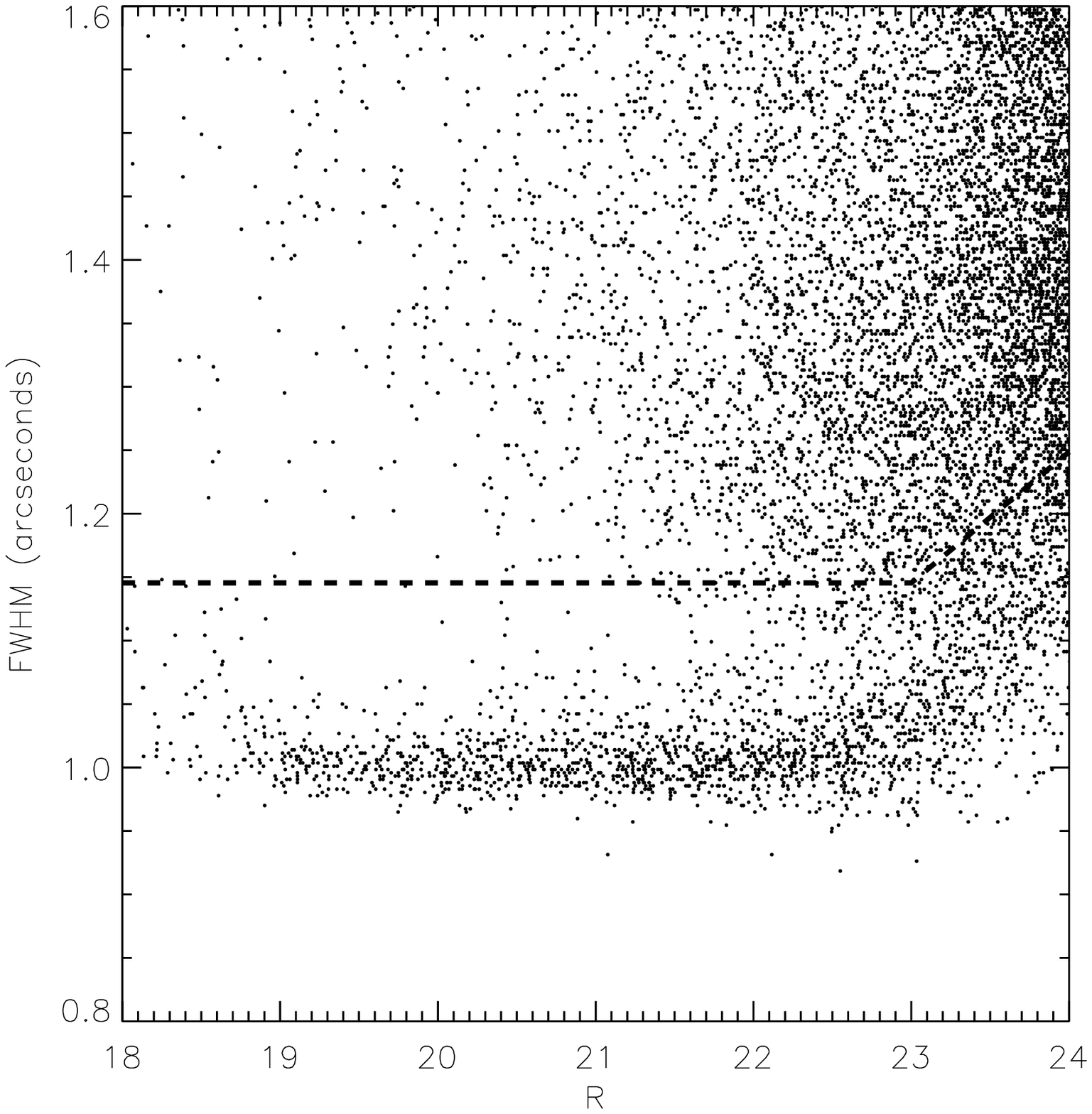}{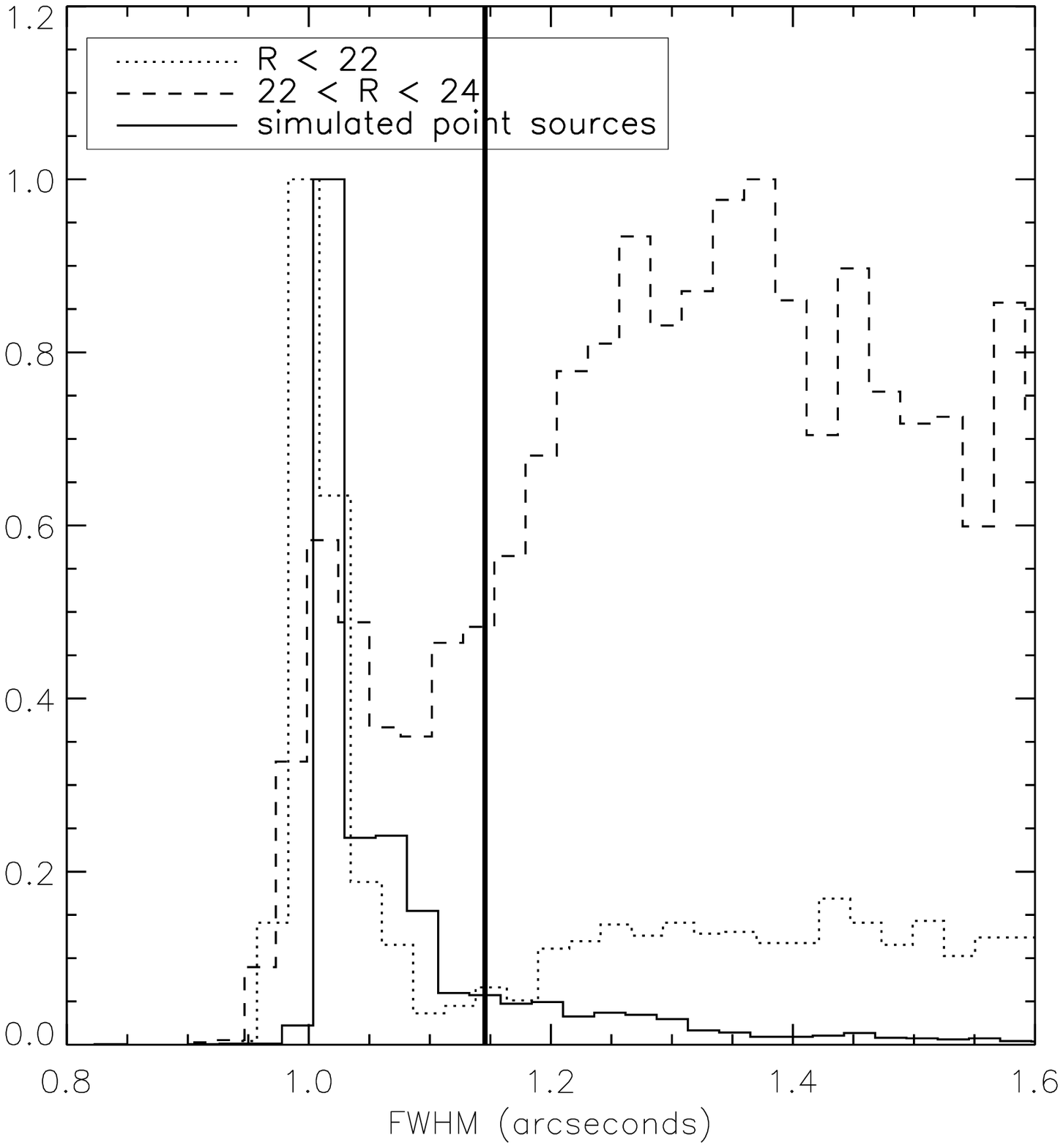}
\caption{Distribution of full width at half-maximum (FWHM) in the $R$-band for objects in NDWFS J1426+3236.  We determine the maximum FWHM for selecting stellar sources.  Left -- FWHM as a function of $R$-magnitude.  The band of stellar sources at FWHM$\sim 1\arcsec$ is obvious.  The dashed line shows the maximum FWHM for our candidate selection. Right --  Normalized histogram of FWHM in two magnitude bins. The dotted line shows objects with $R<22$ and the dashed line shows objects with $22<R<24$. We compare these distributions with the FWHM distribution of simulated stellar sources (solid-line histogram; see text for a details on the simulations) and determine our morphological cutoff at a FWHM that retains $80\%$ of the stellar sources, indicated by the vertical solid line at 1.145\arcsec.  }
\label{fig:fwhm}
\end{figure}

In addition to having the right colors, we required our quasar candidates to be point sources. We use the full width at half-maximum (FWHM) to discriminate between stellar and extended sources.  The left panel of Figure \ref{fig:fwhm} shows the FWHM as measured from the $R$-band image in SExtractor as a function of $R$-magnitude.  A locus of stellar sources with FWHM $\sim 1\arcsec$ is obvious.  

We sought to determine a suitable morphological cut of FWHM that would eliminate galaxy contamination without sacrificing completeness.  To do this, we used the {\tt mkobjects} task in the IRAF {\tt artdata} package to insert 4200 objects with stellar profiles and $17.75 \leq R \leq 24.50$ into each field image.  We re-extract our catalogs using the same SExtractor parameters, and examine the FWHM distribution of our simulated point sources.  This distribution is shown in the right-hand panel of Figure \ref{fig:fwhm} for the same field.  The solid-line histogram shows the distribution of FWHM for simulated point sources, while the dotted line shows the FWHM distribution for sources brighter than $R=22$ and the dashed line shows the FWHM distribution for sources with $22<R<24$.  For each field, we choose as our cutoff the FWHM that retains $80\%$ of our simulated sources.  This cutoff, calculated to be 1.145\arcsec\ for NDWFS J1426+3236, is shown as horizontal dashed line in the left panel of Figure \ref{fig:fwhm} and as a vertical solid line in the right-hand panel.  

\subsection{The Final Candidate List \label{sec:final_cat}}

At $z\gtrsim 4$, neutral hydrogen absorption from the {\lya} forest often causes $B$-band drop outs, even at the deep sensitivities of NDWFS and DLS.  Creating our catalogs in SExtractor's dual-image mode ensures that we measure a flux for each object that has an $R$-band detection, even if an object's flux drops below the $B_W$, $B$ or $V$ detection limits.  Nevertheless, the magnitude assigned to a drop-out object may be unreliable.  

We define the limiting magnitude of each field based on the standard deviation per pixel for a representative region in each image.  We compute the uncertainty in  total flux in a 3\arcsec\ aperture, with area $A$, where each pixel has $1\sigma_{pix}$ counts: $f_{1\sigma} = \sigma_{pix} \sqrt{A} $.  We compute the limiting magnitude from this flux for the $B_W$ fields in NDWFS and the $B$ and $V$ field in DLS.  We assign these limits to any object whose SExtractor-determined magnitude is fainter than these limits.  We then use these limits when applying the color cuts in our candidate selection.  We note that by ignoring the fact that pixel values are correlated in these images we may be underestimating the true noise.  Furthermore, this estimate does not take into account large scale variations from the flat field and other systematic effects.  Nevertheless, this value serves as a practical lower limit on the magnitude in a band without a detection.  Since the limiting magnitudes that we compute are $B, B_W \gtrsim 27$ mags and our color criteria require that $B, B_W - R \geq 2.2$, our survey limit of $R=24$ ensures that all undetected objects meeting our color criteria will be selected.

We ran our catalogs through these selection criteria and ended up with 80 quasar candidates in the five NDWFS fields and 74 candidates in the six DLS fields, strictly based on their cataloged properties.  We then examined image cutouts of the objects and eliminated three candidates from the NDWFS sample and three candidates from the DLS sample that were image artifacts mistaken as objects by SExtractor.  We are therefore left with 77 (16 with $R \leq 23$) NDWFS candidates and 71 DLS (14 with $R \leq 23$) candidates.  We plot these objects as circles in Figure ~\ref{fig:colortracks}, indicating with a right-pointing arrow $B_W$ and $B$-band drop outs.  We list the candidates with $R<23$ in Tables \ref{tab:ndwfs_cands} and \ref{tab:dls_cands} and make the full candidate list available in electronic table format. 

\section{Spectroscopy Results}

We obtained 28 spectra for the DLS and NDWFS candidates and found 23 high-redshift quasars (the remaining objects were three Lyman-break galaxies at $z\sim3.9$, one Type II AGN at $z\sim0.2$ and one featureless, unidentified spectrum).  All but two of our spectra were taken with the Low-Resolution Imaging Spectrometer \citep[LRIS;][]{Oke95} on the Keck I telescope on Mauna Kea.  Five quasars were discovered on UT 2005 November 28 and 29.  Only the red camera was used, with the 400 lines mm$^{-1}$ grating blazed at 8500 \AA.  Sixteen additional quasars were discovered during a run on UT 2006 May 20 to 22.  Again, only the red camera was used, this time with the 600 lines mm$^{-1}$ grating blazed at 7500 \AA.  The spectra were obtained one at a time using the long-slit mode at the parallactic angle.

We reduced the LRIS spectra using BOGUS, an IRAF software package developed by Stern, Bunker, \& Stanford\footnote{Available at \url{http://zwolfkinder.jpl.nasa.gov/~stern/homepage/bogus.html}} to reduce LRIS slitmask data.  We made minor modifications to the code to accommodate the single slit mode of our observations.  BOGUS reduces the two-dimensional spectra, applying a bias-correction, flat-fielding, cosmic ray removal, and sky subtraction.  Most of our spectra were observed in one exposure at a single slit position.  A handful of spectra were dithered in two positions along the slit.  For these objects, BOGUS applied a fringe correction as well.  The {\tt sham\_r} routine (part of the BOGUS package) extracts the one dimensional spectrum and performs wavelength and flux calibrations.  Since the fringing can be quite strong at the position of sky lines at red wavelengths, we interpolated over strong sky lines which appeared in our final spectra using the {\tt skyinterp} IRAF task\footnote{SKYINTERP is part of the WMKONSPEC package for reducing long-slit near-infrared spectra, where imperfect sky subtraction can often produce spectra with strong sky lines (\url{http://www2.keck.hawaii.edu/inst/nirspec/wmkonspec/index.html}). }.

Two additional spectra were taken on UT 2006 March 25 and 26 with 
the Multi-Aperture Red Spectrometer \citep[MARS;][]{Barden01} on
the Mayall 4 meter telescope at Kitt Peak.  MARS is an optical spectrograph
that uses a high resistivity, p-channel Lawrence Berkeley National
Laboratory CCD with little fringing and very high throughput at
long wavelengths.  We obtained spectra of DLS~J105227.9$-$054234
and DLS~J105348.8$-$053319 using the 1\farcs7 wide long slit, OG550
order-sorting filter, and the VG8050 grism.  Across much of the
optical window this instrument configuration delivers resolution
$R \approx 1100$ spectra, as measured from sky lines filling the
slit.  The data were processed following standard optical slit
spectroscopy procedures.  The nights were not photometric, but
relative flux calibration was achieved with observations of
spectrophotometric standards from \citet{Massey90} obtained
during the same observing run.

Figures \ref{fig:spectra} and \ref{fig:spectra2} plot the 23 quasars discovered by our survey, sorted by decreasing redshift.  We mark with a vertical dotted-line expected positions of prominent ultraviolet quasar emission lines: Ly$\alpha$~1216, N~V~1240, Si~IV~1400, C~IV~1550, and C~III]~1909.

\subsection{NDWFS 1433+3408 and DLS 1053$-$0528}

These quasars, at $z=3.88$ and $z=4.02$, respectively, show prominent \ion{N}{4}] $\lambda 1486$, an extremely rare emission line.  When seen in QSOs, this line is normally accompanied by strong \ion{N}{5} $\lambda 1240$ and is a consequence of high nitrogen abundances.  Such objects are dubbed ``nitrogen-rich" and make up at most $0.2\% - 0.7\%$ of QSOs \citep{Osmer80,Baldwin03,Bentz04,Jiang08}.  The QSOs in our survey, however, do not seem to fit into this population.  Their \ion{N}{4}] $\lambda 1486$ equivalent widths (EWs) are orders of magnitude higher than nitrogen-rich QSOs, and their corresponding \ion{N}{5} $\lambda 1240$ is weak, and may even be absorbed in DLS 1053$-$0528 \citep[see][for more details on these objects]{Glikman07a}.

Strong \ion{N}{4}] $\lambda 1486$ emission has been seen in the spectrum of the Lynx arc, a gravitationally lensed \ion{H}{2} galaxy at $z=3.357$ \citep{Fosbury03}.  Here this line can be explained by a hot (80,000 K) blackbody caused by a top-heavy initial mass function (IMF) as the source of ionizing flux -- suggestive of early, metal-poor star formation, e.g., Population III stars. 

\section{Computing the Luminosity Function}

The luminosity function (LF) is defined as the co-moving volume density of objects as a function of luminosity.  In the case of the quasar luminosity function (QLF) it is customary to use the absolute magnitude at 1450\AA, $M_{1450}$.  To construct the QLF for the quasars in our survey, we must first account for the incompleteness of our selection and spectroscopic followup.  We must also compute $M_{1450}$ for each of our quasars.  We describe these steps below and compute the QLF for our survey.  Table \ref{tab:qso_fsel} summarizes our results from the following sections.

\subsection{Survey Completeness \label{sec:fsel}}

To compute the quasar luminosity function, we must be able to correct our quasar counts for incompleteness in our selection technique.  To determine the completeness of our survey as a function of redshift and $R$ magnitude, we follow what is now a standard methodology of simulating the selection probability of quasars in the NDWFS and DLS fields \citep{Warren94,Kennefick95b,Richards06}.  

We begin with our simulated quasar spectra, described in Section \ref{sec:specsim}, in a redshift {\it vs.} $R$-magnitude grid.  The redshift grid points are separated by 0.1 in the range $3.5 < z < 5.2$, and the magnitude grid points are separated by 0.25 magnitudes in the range $17.75 < R < 24.50$.  At each grid-point we simulate 200 spectra generated from a random drawing of distributions of the two power-law slopes, \lya\ emission line EW and one realization of neutral hydrogen intervening absorption systems.  The final set of simulated quasars contained 100,800 spectra that evenly samples our parameter space.  These spectra are convolved with the NDWFS and DLS filter curves to produce tables of magnitudes for all the filters in a given survey, as functions of $R$-magnitude and redshift.

The NDWFS and DLS images have $\sim 45,000$  and $\sim 60,000 - 70,000$ sources detected, respectively, in each of their fields.  We randomly distributed the 100,800 simulated quasars into twenty-four subsets of 4200 quasars, to keep the number of added sources under $10\%$ and avoid over crowding.  To properly insert the simulated quasars into our images as point sources, we determined the shape of the point spread function (PSF) by measuring several hundred well-detected, unsaturated, stars in each image using the IRAF task {\tt imexamine} and determining the $\sigma$-clipped median values of $\alpha$ and $\beta$ in the Moffat profile model,
\begin{equation}
 I = I_c (1 + (r / \alpha)^2)^{-\beta},
\end{equation}
where $I_c$ is the peak value of the intensity, $I$.  We then inserted 4200 quasars into the NDWFS and DLS images, using the IRAF {\tt mkobjects} task in the {\tt artdata} package, in all filters at the same random locations assigning them the appropriate stellar profile parameters, and magnitudes computed from the simulated spectra.  Finally, we added the camera read noise and Poisson noise to the simulated data to closely mimic real sources.

\begin{figure}
\plottwo{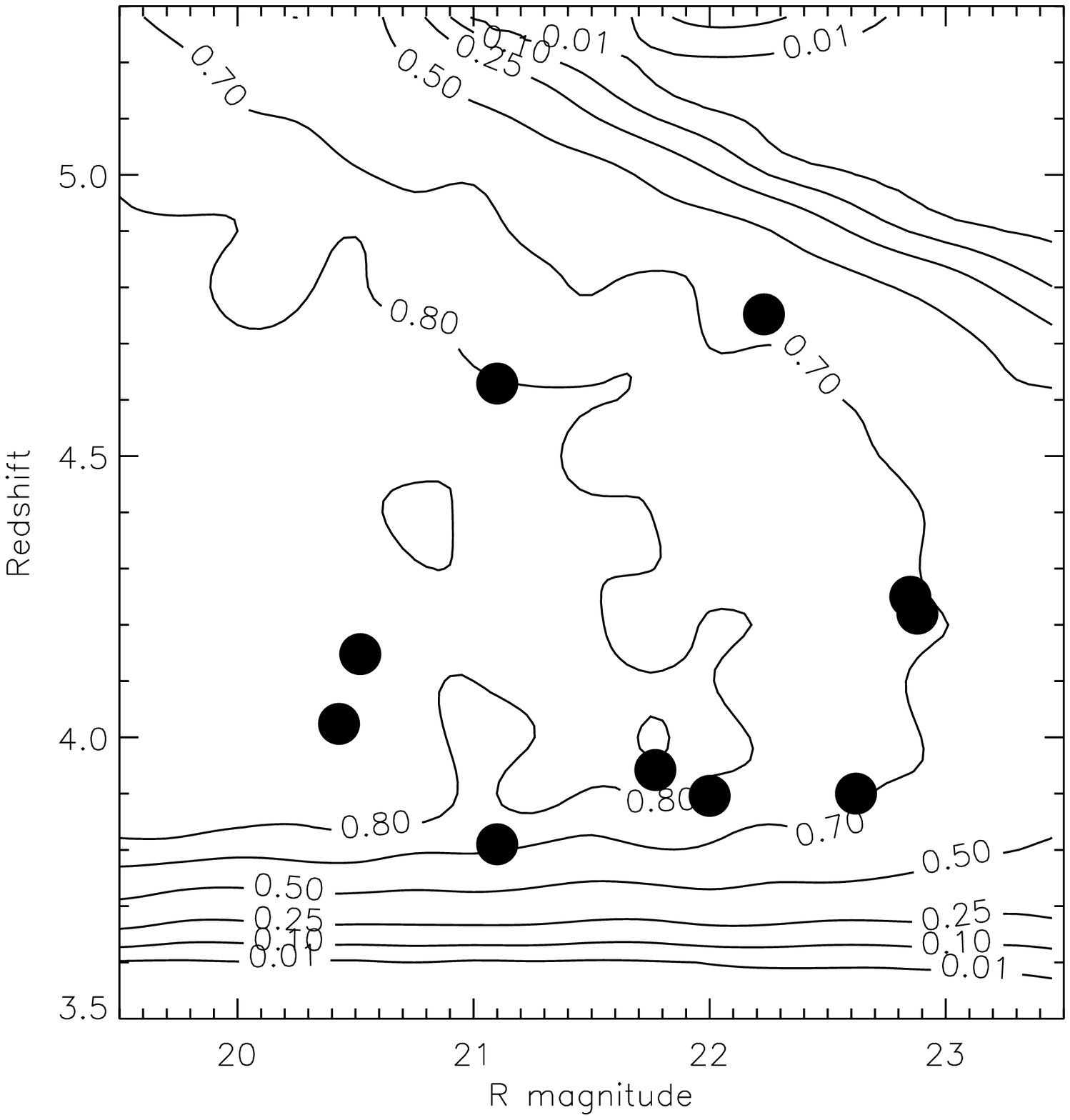}{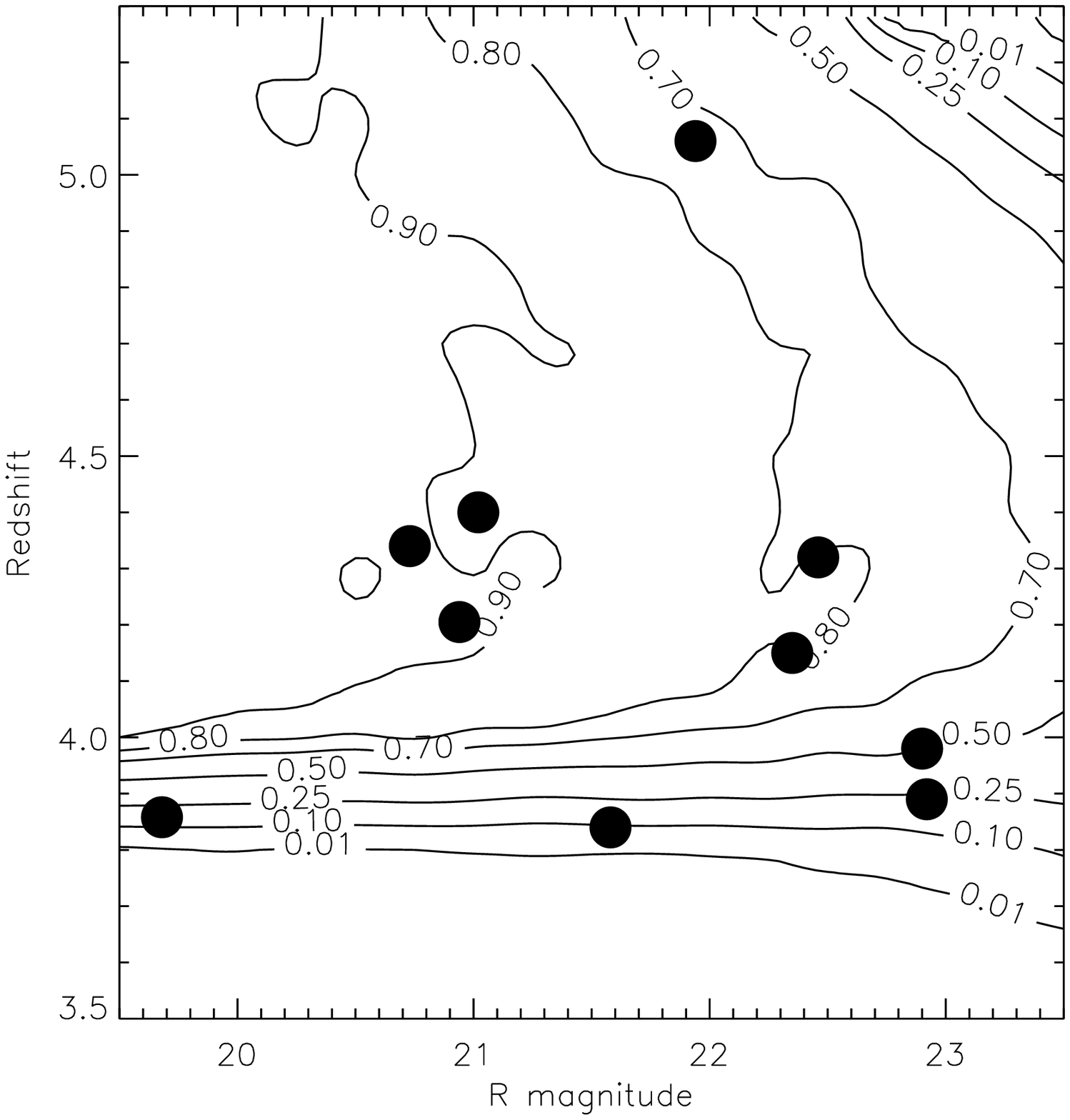}
\caption{Contours representing our survey's completeness fraction for the NDWFS survey on the left and the DLS survey on the right.  Our confirmed quasars are plotted with filled circles.  In most of the magnitude and redshift range of interest ($R \leq 23$, $3.8 \leq z \leq 5.2$), our completeness is above $80\%$, with incompleteness rising at lower redshifts where the \lya\ emission line is just entering the $R$ filter. }
\label{fig:completeness}
\end{figure}

We extracted catalogs for each set of simulated quasars using the identical SExtractor parameters as the unaltered images, using the $R$-band image as the detection frame.  We then processed these objects through our selection pipeline, imposing the color and morphology cuts and assigning magnitude-limits, as described in Section \ref{sec:selection}.  We determine the recovered fraction of quasar candidates in each redshift and magnitude bin (the selection function, $f_{\rm sel}$), thereby mapping the completeness of our candidate selection.  We list the value of $f_{\rm sel}$ in column (4) of Table \ref{tab:qso_fsel}.  Figure \ref{fig:completeness} plots contours showing the completeness as a function of redshift and magnitude for the NDWFS survey on the left and the DLS survey on the right; detected quasars are marked with solid circles. 

Some of our lower redshift quasars exist in areas of high incompleteness, especially in the DLS survey, due to the \lya\ line entering the $R$ band at $z \sim 4$.  Two of these quasars, DLS J0923$+$2952 at $z=3.84$ and DLS J1052$-$0528 at $z=3.89$, have unusually high \lya\ EWs.  Since our simulated quasar \lya\ EW distribution was chosen as a Gaussian distribution with $EW= 65\pm34${\AA}, there were no simulated spectra resembling these objects.  This may point to an underlying problem with simulating quasars using a Gaussian distribution of EWs;  on the other hand these objects may simply be unusual specimens.

\subsection{ Determining K-corrections \label{sec:kcorr} } 

To determine the quasars' absolute luminosity, we need to perform a correction from an observed magnitude to a magnitude at a standard wavelength in the quasar rest-frame.  The absolute magnitude at rest-frame 1450{\AA}, $M_{1450}$, is most commonly used in expressions of QSO luminosity functions.  

Using 200 simulated spectra for each $\Delta z=0.1$ interval in range $3.5 < z < 5.2$  (see Section \ref{sec:specsim}), we calculate the apparent magnitude $m_{1450}$ by convolving each spectrum with a boxcar filter equal to unity in the interval 1425{\AA}$ < \lambda < 1475${\AA} and redshifted to the observed frame. The K-correction is defined as: 

\begin{equation}
 M_{1450} = m - 5\log(D_{l}/10) - K_{\rm corr},
 \label{M1450} 
\end{equation}
where $D_{l}$ is the luminosity distance in parsecs, $M_{1450}$ is the absolute magnitude at rest-frame 1450\AA, and $m$ is the apparent magnitude in some fiducial filter with central wavelength $\lambda$. For a continuum-only quasar spectrum, $F_{\nu} \propto \nu^{\alpha}$, the K-correction to $M_{1450}$ would be equal to:
\begin{equation}
\label{kcorr-nu}
 K_{\rm corr}  = -2.5 (1+\alpha) \log(1+z) - 2.5 \alpha \log(1450\mathrm{\AA}/\lambda),
\end{equation}
where $\lambda$ is the effective wavelength of the filter used for the K-correction. Using simulated quasar spectra, the K-correction can be obtained as the difference between $m$ and apparent magnitude $m_{1450}$ in the observed frame as:
\begin{equation}
 K_{\rm corr}  = m - m_{1450} - 2.5\log(1+z).
\label{Kcorr}
\end{equation}

We use the simulated quasar spectra to determine the expected dispersion in the value of the $K_{\rm corr}$ with redshift. Figure ~\ref{fig:kcorr} shows the value of $K_{\rm corr}$ as well as the one sigma uncertainty interval determined from the various realizations of our simulated quasar spectra. The K-correction for $m_{R}$ (black squares) sharply rises after $z\gsim3.9$, as the {\lya} forest progressively moves into the $R$ band. At our probed redshift range, 1450{\AA} is observed at 6960 - 8990{\AA}.  This suggests that although our selection relies on detections in the $R$ band, our K-corrections can be based on $I$ and $z$-magnitudes for NDWFS and DLS, respectively, minimizing the effect of emission lines on $K_{\rm corr}$.  We plot the corrections to these bands with orange circles and red triangles.  

The blue dashed curve in Figure \ref{fig:kcorr} shows the K-correction for the SDSS $i'$-band magnitude, determined by \citet{Richards06}.  We also convolved our simulated quasar spectra with the SDSS $i'$ filter and obtained the curve plotted with black squares in  Figure ~\ref{fig:kcorr}. The result is extremely similar to the K-correction in \citet{Richards06} below $z=4.8$, after which the inclusion of different descriptions for {\lya} emission and the {\lya} forest simulations in our spectra create a progressive offset.  The $I$ and $z$-band corrections are both consistent with the SDSS corrections out to $z=4.3$ after which they diverge slightly, but never as much as for the $R$-band.  The dispersions in the K-corrections for the SDSS $m_{i}$ and our surveys' magnitudes are equal, having a mean value of $\sigma_{\rm Kcorr}=0.08$ in both cases.  

Because our quasars are at $z \sim 4$ and all our spectra cover 1450\AA\ in the quasars' rest frame, we can also measure $M_{1450}$ directly from the spectra.  To do this, we compute the magnitude from the median flux, $f_{\nu, 1450}$, between rest-frame $\lambda = 1425$\AA\ and $\lambda = 1475$\AA, $m_{1450} = -2.41-2.5\log(f_{\lambda, 1450})$ \citep[Eqn. (2) of][]{Blanton03}.  We also determine the spectrophotometric magnitude, $R_{\rm specphot}$, of each object by convolving its spectrum with the $R$-band filter profile and scaling it to the image-based $R$-band magnitude.  We determine the exact K-correction for each spectrum, independent of any modeling, where 
\begin{equation}
\label{kcorr_spec}
 K_{\rm corr}  = R_{\rm specphot} - m_{1450} - 2.5\log(1+z),
\end{equation}
and apply the K-correction as shown in Equation \ref{M1450}.
 
This in principle should give us a more accurate measure of $M_{1450}$ over the simulated K-corrections.  However, the wavelength range of our LRIS spectra spans 5750-8300\AA, while the $R$ filter profile has $64\%$ transmission at 5750\AA\ and drops to $0\%$ below 5000\AA\ (see Figure \ref{fig:nfilters}).  We could, therefore, be underestimating the flux transmitted through the $R$-band, especially for the lower-redshift objects in our sample.

Figure \ref{fig:absmag} plots the comparison between $M_{1450}$ derived using our simulated spectra (plotted on the vertical axis) and directly from the object spectra (plotted on the horizontal axis).  The triangles are quasars from NDWFS and the asterisks are quasars from DLS.  The dotted line represents a one-to-one correlation, and while the scatter about this line can be significant (up to $\sim 0.3$ magnitude), the best fit line to these points (plotted as a solid black line) has a slope of $1.02\pm0.09$ with an intercept of $0.19\pm2.06$, suggesting an offset of $\sim 0.3$ magnitudes.  The $M_{1450}$ measured directly from the spectra tend to be fainter than $M_{1450}$ estimated from the simulations, which may (at least in part) be because of the incomplete spectroscopic coverage of the $R$-band.

Table \ref{tab:qso_fsel} lists the $M_{1450}$ magnitudes computed from applying the K-corrections from our simulated spectra and from the object spectra directly in columns (6) and (9), respectively.  In the following sections, we use both of these values to compute and analyze the QLF.

\subsection{The $z\sim 4$ Quasar Luminosity Function}

With the selection function and absolute magnitudes in hand, we are able to calculate the QLF at $z\sim 4$.  The $1/V_a$ \citep[also known as the $V/V_{\rm max}$;][]{Schmidt68} method is the most straightforward way to compute the volume density of quasars as a function of luminosity, $\Phi(M_{1450},z=4)$.  The available volume, $V_a$, is the comoving volume within which a quasar with redshift $z$ and magnitude $R$ can be found.  We determine this volume in pencil-beam units of Mpc$^3$ arcmin$^{-2}$ and multiply by the area surveyed, $A$, of the NDWFS and DLS, 1.71  deg$^2$  and 2.05 deg$^2$, respectively, correcting by our selection function and spectroscopic incompleteness (i.e., the fraction of candidates without spectra).

\subsubsection{The $R<23$ Quasars }

For candidates with  $R<23$, our spectroscopic completeness is 0.75 and 0.71 in NDWFS and DLS, respectively.  Beyond this, the completeness drops significantly to 0.33 (2/6) and 0.11 (1/8) for $23 < R \leq 23.5$ (and further, still, for $R > 23.5$ to 0.04 and 0.02) in the NDWFS and DLS, respectively.  This effectively places our survey limit at $R=23$ and we compute here the luminosity function for only these objects.  We divide our sample into one-magnitude-wide bins, which span $-27 < M_{1450} < -20$, for each survey separately, and together, as a combined sample.  
  
While our survey sought to find quasars with $3.8< z \lesssim 5.2$ and brighter than $R=23.0$, corresponding to 4291 Mpc$^3$ arcmin$^{-2}$, this is not always the volume available to the quasars that we find.  For example, a quasar with $R=22.9$ at $z=4.1$ has $M_{1450} = -23.24$.  The limiting redshift for such a quasar would be, $z_{\rm lim}=4.18$, beyond which it would be too faint to make our magnitude limit.  Therefore, $V_a$ for this quasar is the comoving volume between $z=3.8$ to $z_{\rm lim}$, which, in this example, is 1427 Mpc$^3$ arcmin$^{-2}$.  We calculate $z_{\rm lim}$ and $V_a$ for each of our quasars separately, and list them in Table \ref{tab:qso_fsel}.  

The selection function, $f_{\rm sel}$, which we described in \S \ref{sec:fsel}, is the probability that a quasar with magnitude $R$ and redshift $z$ would be selected as a candidate.  For each quasar, we scale $V_a$ by the selection function and spectroscopic completeness, $f_{\rm spec}$ (Column (7) of Table \ref{tab:qso_fsel} referring to the fraction of candidates with spectroscopic followup), in each luminosity bin.  We compute the volume density of quasars for each $M_{1450}$ bin for NDWFS and DLS separately using the following formalism:
\begin{equation}
\Phi(M_{1450},z=4) = \sum_{i}^N (f_{\rm sel}(i)\times f_{\rm spec}(i) \times V_{a}(i) \times A)^{-1}, \label{eqn:phi} 
\end{equation}
where $N$ is the number of quasars in the luminosity bin.  We list the values of $\Phi$ for NDWFS and DLS as well as their average, in Table \ref{tab:lumfunc}.  Figure \ref{fig:lf1} plots $\Phi(M_{1450},z=4)$, using $M_{1450}$ from K-corrected $z$ and $I$ magnitudes on the left and directly from the quasar spectra on the right. We plot the volume densities of NDWFS (squares) and DLS (triangles) quasars as well as for the combined sample (filled circles).   We fit a single-power-law, $\Phi(M) = \Phi^* 10^{-0.4(\beta-1)M}$, to these measurements.  We find that the shape of the QLF depends strongly on how we compute $M_{1450}$.  When we compute $M_{1450}$ using the K-corrected $z$ and $I$ magnitudes, the volume density resembles a power law (with a slope of $\beta = -1.59\pm0.22$) out to the faintest bin, which suffers from the flux-limit bias of our survey (which we discuss in the next subsection).   When we compute $M_{1450}$ directly from the quasar spectra the objects populate fewer bins.  While the shape of this QLF is scattered and does not resemble a power-law, the space densities derived from the individual surveys are more consistent with each other.  Table \ref{tab:splparam} lists our fitted value for $\beta$ from both datasets.

\subsubsection{The $R>23$ Quasars }

Of the 148 candidates that we selected in \S \ref{sec:final_cat}, 117 have $R>23$, 59 in the DLS survey and 63 in the NDWFS survey.  We obtained spectra for six of these objects; two NDWFS sources were Lyman break galaxies (LBGs) at $z\sim 3.8$, three were quasars in our desired redshift range (two in the DLS and one in the NDWFS) and one NDWFS source was a Type-II AGN at $z=0.21$.  We wish to make use of the three $R>23$ quasars to enhance our measurement of the QLF.

We compute the volume density for these quasars according to Equation \ref{eqn:phi}.  However, we determine the limiting redshift, $z_{\rm lim}$, and $V_a$ assuming a survey magnitude limit of $R=23.5$ and $R=24$, when appropriate.  The largest source of uncertainty in this method arrives from spectroscopic completeness, $f_{\rm spec}$, which we use to scale our available volume $V_{a}$ and which suffers from small number statistics.  For example, in the DLS sample, we obtained one spectrum (of a quasar) out of nine candidates with $23 < R \leq 23.5$, rendering $f_{\rm spec} = \frac{1}{9} = 0.11$ and the Poisson error on this is $0.12$. 

These values are listed in Table \ref{tab:qso_fsel} and we compute the QLF with these three additional quasars using both estimates of $M_{1450}$.  We plot the QLF including all of our quasars in Figure \ref{fig:lf2}.  Again, we fit a single power-law to these data and find that regardless of how we compute $M_{1450}$, the QLF resembles a power-law.  The slopes are marginally consistent, and steeper than the fit using only quasars with $R \leq 23$.  This is not surprising, as the volume densities for the faintest quasars are potentially significantly inflated because of the low spectroscopic completeness, which appears in the denominator of Equation \ref{eqn:phi}.  Nevertheless, including these data appear to remedy the divergent volume densities computed for NDWFS and DLS in the left-hand panel of Figure \ref{fig:lf1} as well as the scattered shape of the QLF in the right-hand panel of Figure \ref{fig:lf1}.

We list the volume density for our QLF including all 23 quasars for both methods of computing $M_{1450}$ in Table \ref{tab:lumfunc} and the best-fit slope in Table \ref{tab:splparam}.

\subsection{Comparison with other surveys}

We compare our binned QLF data points for $3.8 < z < 5.2$ (median $z=4.15$) with the results from other surveys at similar redshifts in Figure \ref{fig:qlf_all}.  Our results for the binned QLF using all the quasars in our survey, including those with $R>23$, are plotted with red circles.  The left-hand panel shows our QLF where $M_{1450}$ has been computed from K-corrections to the $z$ and $I$ band photometry and the right-hand panel shows the QLF where $M_{1450}$ was obtained directly from the spectra.  We plot with a dotted line the best-fit power-law for the faint end, shown in Figure \ref{fig:lf2}.

To extend our QLF into the bright end, the most suitable sample comes from the SDSS.  \citet{Richards06} determined the QLF for SDSS quasars from the Third Data Release (DR3).  We utilize their results for the $z=4.25$ bin and plot their binned LF with asterisks and their single-power-law model fit with a dash-dot-dot-dot line in Figure \ref{fig:qlf_all}.  Both the points and the curve are evolved to $z=4.15$ using the evolution model determined by  \citet{Richards06}. 

\citet{Fontanot07a} present a re-analysis of the SDSS-DR3 quasars with $M_{1450} < -26$ combined with eleven quasars from the {\it Great Observatories Origins Deep Survey} (GOODS) with $M_{1450} > -23.5 $ in the range $3.5 < z < 5.2$.  We plot their QLF at $z=3.75$ and $z=4.6$ with open triangles and squares in Figure \ref{fig:qlf_all}.   We evolve both these data sets to $z=4.15$ using the same evolution model as above.  The pure density evolution (PDE) model fit from \citet{Fontanot07a}, also evolved to $z=4.15$, is plotted with a dash-dot line.

While \citet{Richards06} and \citet{Fontanot07a} rely on the same SDSS-DR3 quasars, their derived volume densities are highly discrepant.  The difference between the LFs stems from their different estimates for the completeness of the SDSS sample as a function of redshift.   \citet{Fontanot07a} estimate the completeness for the SDSS subsample used in their calculation by building a spectral library based on the quasar template spectrum from \citet{Cristiani90}.  Instead of modifying the composite using a Gaussian distribution of power-law continua \citep[as is done in this work and in][]{Richards03}, they empirically determine a distribution of continua blueward of Ly$\alpha$ from 215 high quality quasar spectra with $2.2 < z < 2.25$ in the SDSS-DR3 quasar sample. As a consequence, the distribution of power-law slopes in the \citet{Fontanot07a} library is steeper, with $\alpha_2 = -0.7 \pm 0.3$, compared with $\alpha_2 = -0.5\pm0.3$ in the \citet{Richards06} treatment.  The IGM absorption is then added in a manner similar to ours.  They use this spectral library to estimate the completeness of the SDSS color selection criteria \citep{Richards02} and find that the completeness drops to 50\% (and as low as 30\% at $z=4.4$) for $3.6 < z < 4.4$.  At $4.5<z<4.9$ the completeness rises to above 90\% and drops off again after \citep[see][Figure A.4.]{Fontanot07a}\footnote{This is discussed in detail in appendix (A.3) of \citet{Fontanot07a}.}

Using a method similar to ours to create their spectral library, \citet{Richards06} calculate a completeness of nearly $90\%$ for $z > 3.8$.  While our method more closely matches that of  \citet{Richards06}, our values seem to have better continuity with \citet{Fontanot07a}, especially for the QLF whose $M_{1450}$ was derived from K-corrected $z$ and $I$ band magnitudes (the left-hand panel of Figure \ref{fig:qlf_all}).  

For comparison, we overplot in Figure \ref{fig:qlf_all} QLF model fits for several surveys at or near our survey's redshift.  The left-hand panel shows general agreement between our QLF based on $M_{1450}$ computed from simulated quasars (in all but the faintest bins) and the SDSS+GOODS analysis of \citet{Fontanot07a} as well as the VVDS survey \citep{Bongiorno07}.  The latter spectroscopically identified 130 broad-line AGN with $z=0-5$ over $\sim 1.7$ deg$^2$ in two fields.  The $z=3.0-4.0$ bin from their QLF contains 17 QSOs with a luminosity range of $-26 \lesssim M_{1450} \lesssim -22$ and is well-suited for comparison with our data.  We plot their model fit with a solid black line in Figure \ref{fig:qlf_all}.  

There is a large discrepancy between our measurements and predictions for the value of the quasar density at $z\sim 4$.  The COMBO-17 survey \citep{Wolf03} covered 0.78 deg$^2$ to $R<24$ and found 11 QSOs in the redshift range $3.6 < z < 4.2$, spanning the luminosity range $-27 < M_{1450} < -24$ and 4 QSOs in the redshift range $4.3 < z < 4.8$ in the luminosity range of $-27 < M_{1450} < -26$.  While their results are in agreement with SDSS at the bright end, their QLF modeled as Pure Density Evolution (PDE, dashed line in Figure \ref{fig:qlf_all}) significantly underpredicts the number of quasars at the faint end.

\subsection{Model Fitting to the Combined QLF}

The shape of the QLF is typically parametrized by a standard double power-law form:
\begin{equation}
\Phi(M,z) = \frac{\Phi(M^*)}{10^{0.4(\alpha+1)(M-M^*)}+10^{0.4(\beta+1)(M-M^*)}}. \label{eqn:dpl} 
\end{equation}
We follow the formalism of \citet{Sandage79} and \citet{Marchesini07} to determine the best fit parameter values for the QLF using the maximum likelihood (ML) method.  We combine the SDSS quasars at $z\sim 4$ with our 23 quasars to sample as broad a magnitude range as possible when fitting the double-power-law in equation \ref{eqn:dpl}.   We try different combinations of our QLF, based on the two ways of calculating $M_{1450}$, with the QLF from  \citet{Richards06} and \citet{Fontanot07a} at the bright end.  As is clear from Figure \ref{fig:qlf_all}, the bright-end QLF from \citet{Fontanot07a} has better continuity with our QLF based on $M_{1450}$ computed from simulated quasars.  There is a sharp jump between our values and  \citet{Richards06}.  We therefore do not attempt to fit these points in this case.  We do not attempt to model the evolution of the QLF, since our survey is focused on a ``snapshot" of the quasar population at a narrow redshift range.  

We initially attempted to maximize the likelihood $\Lambda = \prod^N_{i=1} p_i$ with respect to all three parameters, $\alpha, \beta$ and $M^*$.  We created a gridded cube of $\alpha$, $\beta$ and $M^*$ values, spanning $\alpha,\beta = [-4,0]$ and $M^*=[-27,-20]$, and computed $\Lambda$ at each point, first using a ``coarse" grid ($\Delta\alpha,\beta = 0.1$, $\Delta M^* = 0.2$) with the intention of refining the values near the maximum.  However, due to the small number of quasars in our sample, poorly constraining the faint end, no clear maximum could be found over these reasonable ranges of $\alpha$, $\beta$ and $M^*$.

Since the ML method fails to constrain the double-power-law parameters, we fit Equation \ref{eqn:dpl} to the binned QLF using a $\chi^2$ minimization and allowing all four parameters, $\Phi^*$, $M^*$, $\alpha$, and $\beta$, to vary.   The results of our fit are listed in Table \ref{tab:dplparam} for the three combinations of data sets that we fit.  We plot with a red line in Figure \ref{fig:qlf_all} the best-fit double-power-law parameters from dataset (1) in the left-hand panel and dataset (3) on the right-hand panel.

When our QLF is combined with SDSS data on the bright end, the best-fit parameters are more strongly influenced by the choice of bright-end dataset than our method for estimating $M_{1450}$.  This is because the error bars on the bright-end bins are much smaller compared to the faint-end and strongly constrain the fit.  Furthermore, despite the different shapes of the QLF in our two derivations, the fit using the \citet{Fontanot07a} are nearly identical and largely unaffected by the four bins made up of GOODS quasars on the faint end; we excluded these bins and obtained nearly the same values for the double-power-law parameters. 

 When fitting to our data combined with \citet{Richards06}, we compute, effectively, a single power law fit, with a slope $\alpha \simeq \beta = -2.3\pm0.2$.  This result is provocative, both because it preserves the slopes measured for each data set individually \citep[$\alpha \sim 2.1$ for $z=4.25$ in Figure 21 of][while we find $\beta = -2.46$ in the right-hand panel of Figure \ref{fig:lf2}]{Richards06}, and because this suggests that the number of faint AGN rises dramatically out to very faint magnitudes, and that there is no observed ``knee" to the QLF at $z\sim 4$.  This degeneracy between $\alpha$, $\beta$ and $M^*$ explains why the ML method for the double-power-law fit did not converge. Therefore, we fit a single power-law, $\Phi(M) = \Phi^* 10^{-0.4(\beta-1)M}$, to this data set using the ML method.  In this case there is one free parameter, $\beta$, which we allow to vary between $-1$ and $-4$ in steps of $\Delta\beta=0.01$.  The likelihood is maximized at $\beta = -2.67$ and the $68\%$ confidence limits are $(-2.80,-2.55)$, the $90\%$ confidence limits are $(-2.88,-2.48)$ and the $99\%$ confidence limits are $(-3.00,-2.37)$.  While this result stands in conflict with other measurements of the QLF to faint magnitudes which see a flattening of the faint end with redshift \citep[e.g.,][]{Wolf03,Hunt04}, we discuss its plausibility and possible interpretation in \S \ref{sec:discussion}.

\section{Estimating the Contribution of Quasars to the UV Radiation Field at $z\sim 4$}

We use the best-fit parameters from our $\chi^2$ minimization to all the binned data to integrate Equation \ref{eqn:dpl} and compute the emissivity of quasars at 1450\AA, $\epsilon_{1450}$.  Our computed UV radiation depends strongly on the best-fit parameters (listed in Table \ref{tab:dplparam}) for the QLF, especially the faint-end slope, $\beta$.  

For dataset (1), where we combined our binned QLF using K-corrected $z$ and $I$-band photometry with the data from \citet{Fontanot07a}, we compute  $\epsilon_{1450} \cong 9 \times 10^{25}$ erg s$^{-1}$ Hz$^{-1}$ Mpc$^{-3}$.  Following \citet{Madau99}, we use our parametrized QLF together with the quasar SED of \citet{Elvis94} to compute the photoionization rate from QSOs at this redshift, $\dot{N}_{\rm QSO}  = 4 \times 10^{51}$ s$^{-1}$ (integrating $\int \Phi(L)L dL$ down to $M_{1450} = -20$).  This is almost twice the value needed to ionize the intergalactic medium (IGM) at $z = 4.15$, $\dot{N}_{\rm IGM} = 2.4 \times 10^{51}$ s$^{-1}$ and stands in contrast to previous statements on the ability of quasars to ionize the universe at higher redshifts \citep{Haiman01,Wyithe03a,Shankar07} which suggested that AGN do not produce a sufficient number of ionizing photons and starforming galaxies must play a larger role.

For dataset (3), where we combined our binned QLF using $M_{1450}$ from the object spectra with the data from \citet{Richards06} and effectively fit a single power law, we find $\epsilon_{1450}=7 \times 10^{25}$ erg s$^{-1}$ Hz$^{-1}$ Mpc$^{-3}$ and $\dot{N}_{QSO} = 3 \times 10^{51}$ photons s$^{-1}$ Mpc$^{-3}$, which also produces sufficient photons to ionize the IGM.  This value is strongly dependent on the faint limit of integration.  Because the total UV-luminosity density has to be larger from the contribution of quasars fainter than our integration limit, our computation of $\dot{N}_{QSO}$ is a lower limit of the photoionization rate. 

\section{Discussion \label{sec:discussion}}

Our survey has found that the comoving volume density of quasars continues to rise as a steep power law to low luminosities, four magnitudes fainter than previous measurements.  This is true regardless of how $M_{1450}$ is computed.  This result conflicts with predictions for the evolution of the shape of the QLF with redshift, based on the observed evolution of the QLF at lower $z$.  In addition, this is surprising because objects at the faint end of a flux-limited survey, with their associated large photometric errors, tend to be undercounted.  While our results are sensitive to the fluctuations of small number statistics (our faintest bin contains between one and two objects, depending on how $M_{1450}$ is computed) it behooves us to come up with a reasonable explanation or rule out any systematics that may lead to an excess of quasars counted in error.  

In this section, we consider possible explanations for overcounting our quasars at the faintest bins.  These are: (1) we are including galaxies in our sample, (2) our selection function overcorrects $\Phi(M)$ (equation \ref{eqn:phi}) leading to an overestimate of the QLF, (3) clustering of quasars in our chosen fields due to large scale structure.  We examine each of these possibilities below.

We rule out possibility (1), that we have counted galaxy interlopers as quasars, by an examination of Figure \ref{fig:spectra}.  All of the spectra except for our lowest-redshift object, NDWFS J142713.2+322842 at $z=3.74$, show strong, broad Lyman-$\alpha$ emission with the classic asymmetric profile from absorption of the blue wing of the emission line attributed to high-redshift quasars. NDWFS J142713.2+322842 is also obviously a quasar, as its \ion{C}{4} line is extremely broad, with FWHM $\simeq 11,000$ km s$^{-1}$.

To examine possibility (2), that our number counts are too high by an underestimate of the selection function, we plot the value of the selection function for each quasar versus its absolute magnitude, $M_{1450}$ in Figure \ref{fig:fselvm1450}.  As we already noted in Section \ref{sec:fsel}, $f_{\rm sel}$ is $\sim 80\%$ for most of the quasars.  At faint luminosities, $f_{\rm sel}$ drops to $\sim 0.4$, which means that $\Phi(M_{1450},z=4)$ in equation \ref{eqn:phi} is corrected for incompleteness by a factor of $\sim 2.5$.  This is insufficient to account for the steep rise in $\Phi(M)$ that we see, since a flattening of the faint-end slope of the QLF would predict fewer quasars by a factor $\gtrsim 50$. 

The third possibility is that clustering due to  cosmic variance is enhancing the number of quasars that we find, which can have a significant impact in a small sample.  We examined the spatial and redshift distribution of our quasars for each field and found that of the three faintest and lowest luminosity quasars (which dominate the last two bins in our QLF)  are isolated, with the nearest quasar $\gtrsim$ 0.5 deg away.  This corresponds to a separation on the sky of $\sim$12 Mpc at $z\sim4.15$.

 DLS J105346.1$-$052859, the faintest quasar in our sample with $R=23.83$, is 4\farcm4 away from DLS J105348.8$-$053319, a much brighter quasar with $R=20.94$.  While these objects are the closest pair of quasars in our sample, their redshift difference (the faint quasar is at $z=4.02$ and the brighter quasar is at $z=4.20$, implying that their orthogonal separation is $\sim 2000$ Mpc in our stated cosmology) rules that their proximity is merely a projection effect.  

There are a few quasar samples that, while small, are consistent with steeper QLF at fainter luminosities.  \citet{Cool06} found three $z> 5$ quasars in NDWFS Bo\"{o}tes using mid-infrared color selection, which is meant to be insensitive to dust reddening and avoids confusion with the stellar locus \citep{Stern05}.  None of these quasars appear in our candidate list from NDWFS.  This is because our color selection becomes highly incomplete at  $z>5$ and $R\gtrsim 23$.  In addition, the optical colors of the quasars from \citet{Cool06} fail to meet our color criteria.  Only one object, J142937.9+330416, has $B_W-R>2.2$, and none of them met the criteria set forth in eqn. (2).  While their objects are somewhat more luminous than our $z\sim 4$ quasars, with $M_{1450} = -26$ to $-24.5$, they are $1.5$ magnitudes fainter than the SDSS quasars at these redshifts.  The space density derived from the discovery of these three quasars is consistent with a luminosity function that has a steep power-law index, $\Phi(L) \propto L^{-3.2}$.  The discovery of a radio-loud, $z=6.12$ quasar in NDWFS (the probability of which is $< 1\%$) by \citet{McGreer06} also suggests that the space density of faint, high-redshift quasars may be higher than previously thought.

\section{Conclusion}

Using 23 quasars at $z\sim 4$ discovered in our survey of deep optical imaging data from the DLS and NDWFS surveys, we have measured the faint end of the QLF.  Depending on how we compute $M_{1450}$, our directly fit faint-end slope ranges between $\beta = -1.98\pm0.23$ and $\beta=-2.46\pm-0.20$, in both cases steeper than the faint-end slope measured by \citet{Fontanot07a} who find $\beta = -1.71\pm0.41$.  When we combine our QLF at the faint end with with the SDSS-based QLF at the bright end from \citet{Richards06} we conclude that the shape of the $z\sim 4$ QLF is best fit by a single power law, with a slope of $\sim -2.7\pm0.1$ (via the ML method). A QLF with this shape is able to produce enough UV photons to ionize the IGM at this redshift, which is unexpected.

This result is provocative and presents a challenge to interpretations of the shape of the QLF and its evolution \citep[e.g.][]{Hopkins06b,Shankar07} and has two immediate cosmological implications: (1) the models of the faint AGN evolution
at the epochs $\sim1$ Gyr after the end of the reionization, and possibly all the way into the reionization era, may need to be revisited; (2) AGN were a more significant contributor to the metagalactic ionizing UV flux at these epochs, affecting the evolution of the IGM.  We caution, however, that this result is currently reliant on only a few quasars at the faintest end of our survey.  Additional spectra of our $R>23$ candidates will provide a more robust measurement in this crucial luminosity regime and better constraints for theoretical models.\\ \\

\acknowledgments
We are grateful to the staff of W. M. Keck observatory for their assistance during our observing runs. This work was supported in part by the NSF grants AST-0407448 and AST-0909182, and by the Ajax foundation.  The work of DS was carried out at Jet Propulsion Laboratory, California Institute of Technology, under a contract with NASA.

This work makes use of image data from the NOAO Deep Wide-Field Survey (NDWFS) and the Deep Lens Survey (DLS) as distributed by the NOAO Science Archive. NOAO is operated by the Association of Universities for Research in Astronomy (AURA), Inc. under a cooperative agreement with the National Science Foundation.

\pagebreak


\begin{deluxetable}{cccccccccc}

\rotate

\tablewidth{0pt}


    \tablecaption{NDWFS Fields Surveyed \label{tabl:ndwfs_fields}}


\tablehead{\colhead{Field} & \colhead{R.A.} & \colhead{Dec} & \colhead{Size} & \colhead{FWHM$_{BW}$} & \colhead{FWHM$_R$} & \colhead{FWHM$_I$} & \colhead{EXP$_{BW}$} & \colhead{EXP$_R$} & \colhead{EXP$_I$} \\
\colhead{} & \colhead{(J2000)} & \colhead{(J2000)} & \colhead{(\arcmin)} & \colhead{(\arcsec)} & \colhead{(\arcsec)} & \colhead{(\arcsec)} & \colhead{(sec)} & \colhead{(sec)} & \colhead{(sec)} } 

\startdata
NDWFS J1426$+$3236  &  14:26:03.74  &  $+$32:36:31.72  &  $36.8\times38.0$  &  0.93  &  0.94  &  0.96  &  8400  &   6000  &  12000 \\
NDWFS J1431$+$3236  &  14:31:36.14  &  $+$32:36:46.29  &  $36.8\times38.0$  &  0.80  &  0.91  &  0.77  &  8400  &   6000  &  12000 \\
NDWFS J1434$+$3421  &  14:34:30.79  &  $+$34:21:54.18  &  $35.8\times37.4$  &  1.05  &  0.98  &  0.79  &  8400  &   4200  &  10200 \\
NDWFS J1437$+$3347  &  14:37:16.32  &  $+$33:47:01.72  &  $36.9\times38.0$  &  0.89  &  0.87  &  0.88  &  8400  &   6000  &  12000 \\
NDWFS J1437$+$3457  &  14:37:24.59  &  $+$34:57:02.13  &  $35.5\times38.0$  &  0.86  &  1.07  &  1.16  &  8400  &  10800  &  11400 \\
\enddata




\end{deluxetable}


\begin{deluxetable}{cccccccccc}

\rotate

\tablewidth{0pt}


\tablecaption{DLS Fields Surveyed\label{tab:dls_fields}}


\tablehead{\colhead{Field} & \colhead{R.A.} & \colhead{Dec} & \colhead{Size} & \colhead{FWHM$_B$} & \colhead{FWHM$_V$} & \colhead{FWHM$_R$} & \colhead{FWHM$_z$} & \colhead{EXP$_R$} & \colhead{EXP$_BVz$} \\ 
\colhead{} & \colhead{(J2000)} & \colhead{(J2000)} & \colhead{(\arcmin)} & \colhead{(\arcsec)} & \colhead{(\arcsec)} & \colhead{(\arcsec)} & \colhead{(\arcsec)} & \colhead{(sec)} & \colhead{(sec)} } 

\startdata
   F2p21  &  09:22:37.1   & $+$30:00:00  &  $35.1\times 35.1$  &  1.34  &  1.08  &  0.86  &  0.88  &  18000  &  12000 \\
   F2p31  &  09:22:37.1   & $+$29:20:00  &  $35.1\times 35.1$  &  1.26  &  1.05  &  0.78  &  1.50  &  18000  &  12000 \\
   F2p33  &  09:16:27.7   & $+$29:20:00  &  $35.1\times 35.1$  &  1.24  &  1.02  &  0.86  &  1.28  &  18000  &  12000 \\
   F4p23  &  10:49:19.4   & $-$05:00:00  &  $35.1\times 35.1$  &  1.09  &  1.10  &  0.89  &  1.12  &  18000  &  12000 \\
   F4p31  &  10:54:40.8   & $-$05:40:00  &  $35.1\times 35.1$  &  1.20  &  0.99  &  0.87  &  1.11  &  18000  &  12000 \\
   F4p32  &  10:52:00.0   & $-$05:40:00  &  $35.1\times 35.1$  &  1.17  &  0.93  &  0.89  &  1.24  &  18000  &  12000 \\
\enddata




\end{deluxetable}


\begin{deluxetable}{ccccccccccc}

\rotate

\tabletypesize{\scriptsize}

\tablewidth{0pt}

\tablecaption{NDWFS Candidates \label{tab:ndwfs_cands}}


\tablehead{\colhead{R.A.} & \colhead{Dec} & \colhead{Field} & \colhead{$B_W$} & \colhead{$R$} & \colhead{$I$} & \colhead{$B_w-R$} & \colhead{$R-I$} & \colhead{FWHM$_R$} &  \colhead{Redshift} & \colhead{Class.} \\ 
\colhead{(J2000)} & \colhead{(J2000)} & \colhead{} & \colhead{(mag)} & \colhead{(mag)} & \colhead{(mag)} & \colhead{(mag)} & \colhead{(mag)} & \colhead{(arcsec)}  & \colhead{} & \colhead{} } 

\startdata
14 25 22.71 &  $+$32 48 27.0 &  1426p3236 &  23.35$\pm$0.02 &  21.10$\pm$0.01 &  20.96$\pm$0.01 &   2.25 &     0.14 &   1.08 &    3.81    &  QSO\\ 
14 30 17.78 &  $+$32 20 03.4 &  1431p3236 &  $>$27.90       &  22.23$\pm$0.01 &  21.43$\pm$0.01 &   5.68 &     0.80 &   1.03 &    4.75    &  QSO\\ 
14 30 35.22 &  $+$32 44 26.3 &  1431p3236 &  24.08$\pm$0.03 &  21.49$\pm$0.01 &  21.50$\pm$0.01 &   2.59 &  $-$0.01 &   0.99 &    0.21    &  AGN II\\ 
14 31 01.94 &  $+$32 51 55.5 &  1431p3236 &  25.52$\pm$0.11 &  21.77$\pm$0.01 &  21.17$\pm$0.01 &   3.75 &     0.60 &   0.97 &    3.94    &  QSO\\ 
14 31 52.32 &  $+$32 54 50.2 &  1431p3236 &  23.94$\pm$0.04 &  21.42$\pm$0.01 &  21.12$\pm$0.01 &   2.51 &     0.30 &   1.03 &    \nodata &  \nodata\\ 
14 32 04.74 &  $+$32 54 05.8 &  1431p3236 &  24.61$\pm$0.05 &  22.31$\pm$0.01 &  22.10$\pm$0.01 &   2.30 &     0.21 &   0.97 &    \nodata &  \nodata\\ 
14 33 24.54 &  $+$34 08 41.2 &  1434p3421 &  26.13$\pm$0.19 &  22.62$\pm$0.02 &  22.61$\pm$0.02 &   3.51 &     0.01 &   0.92 &    3.88    &  QSO\\ 
14 33 31.15 &  $+$34 32 48.3 &  1434p3421 &  24.24$\pm$0.04 &  20.52$\pm$0.00 &  20.53$\pm$0.00 &   3.71 &     0.00 &   0.92 &    4.15    &  QSO\\ 
14 36 26.69 &  $+$34 49 50.1 &  1437p3457 &  25.38$\pm$0.10 &  22.93$\pm$0.02 &  22.73$\pm$0.03 &   2.45 &     0.20 &   1.13 &    \nodata &  \nodata\\ 
14 36 42.86 &  $+$35 09 23.8 &  1437p3457 &  24.66$\pm$0.05 &  22.00$\pm$0.01 &  21.95$\pm$0.02 &   2.66 &     0.05 &   1.17 &    3.90    &  QSO \\ 
14 36 58.34 &  $+$33 36 32.0 &  1437p3347 &  23.72$\pm$0.02 &  20.43$\pm$0.00 &  20.31$\pm$0.00 &   3.29 &     0.12 &   0.93 &    4.02    &  QSO\\ 
14 37 32.67 &  $+$33 55 22.0 &  1437p3347 &  26.61$\pm$0.35 &  22.88$\pm$0.02 &  22.57$\pm$0.03 &   3.74 &     0.30 &   0.96 &    4.22    &  QSO\\ 
14 37 34.26 &  $+$34 53 32.9 &  1437p3457 &  25.39$\pm$0.10 &  22.65$\pm$0.02 &  22.33$\pm$0.02 &   2.74 &     0.33 &   1.14 &    \nodata &  ?\\ 
14 37 50.50 &  $+$34 59 52.9 &  1437p3457 &  25.49$\pm$0.11 &  22.94$\pm$0.02 &  22.63$\pm$0.03 &   2.55 &     0.31 &   1.27 &    \nodata &  \nodata\\ 
14 38 13.85 &  $+$35 02 36.4 &  1437p3457 &  25.75$\pm$0.13 &  22.85$\pm$0.02 &  22.67$\pm$0.03 &   2.90 &     0.18 &   1.26 &    4.25    &  QSO\\ 
14 38 39.68 &  $+$35 12 45.9 &  1437p3457 &  24.69$\pm$0.05 &  21.10$\pm$0.00 &  20.52$\pm$0.01 &   3.59 &     0.58 &   1.12 &    4.63    &  QSO\\ 
\enddata




\end{deluxetable}


\begin{deluxetable}{ccccccccccccc}

\rotate

\tabletypesize{\scriptsize}

\tablewidth{0pt}

\tablecaption{DLS Candidates \label{tab:dls_cands}}


\tablehead{\colhead{R.A.} & \colhead{Dec} & \colhead{Field} & \colhead{$B$} & \colhead{$V$} & \colhead{$R$} & \colhead{$z$} & \colhead{$B-R$} & \colhead{$B-V$} & \colhead{$R-z$} & \colhead{FWHM$_R$} &  \colhead{Redshift} & \colhead{Class.} \\ 
\colhead{(J2000)} & \colhead{(J2000)} & \colhead{} & \colhead{(mag)} & \colhead{(mag)} & \colhead{(mag)} & \colhead{(mag)} & \colhead{(mag)} & \colhead{(mag)} & \colhead{(mag)} & \colhead{} & \colhead{} &  \colhead{} } 

\startdata
09 15 27.53  & $+$29 17 50.4  & F2p33  & 24.73$\pm$0.07          & 21.76$\pm$0.01  & 20.72$\pm$0.00  & 20.16$\pm$0.01  &  4.02  &  2.97  &    0.55  & 0.89   & 4.34     & QSO\\ 
09 21 24.62  & $+$29 53 30.8  & F2p21  & 24.91$\pm$0.09          & 23.42$\pm$0.03  & 22.57$\pm$0.01  & 23.28$\pm$0.13  &  2.34  &  1.49  & $-$0.71  & 0.87   & \nodata  & \nodata  \\ 
09 21 51.96  & $+$29 24 57.2  & F2p31  & 25.41$\pm$0.14          & 23.44$\pm$0.03  & 22.46$\pm$0.01  & 21.86$\pm$0.04  &  2.95  &  1.97  &    0.60  & 0.81   &  4.32     & QSO\\ 
09 22 36.49  & $+$30 10 10.6  & F2p21  & 25.53$\pm$0.16          & 24.23$\pm$0.07  & 22.90$\pm$0.01  & 22.94$\pm$0.10  &  2.63  &  1.30  & $-$0.04  & 0.88   &  3.98     & QSO\\ 
09 23 27.22  & $+$29 52 51.7  & F2p21  & 23.82$\pm$0.04          & 22.28$\pm$0.01  & 21.58$\pm$0.01  & 21.66$\pm$0.03  &  2.23  &  1.54  & $-$0.08  & 0.97   & 3.84     & QSO\\ 
09 23 36.82  & $+$30 09 49.9  & F2p21  & $>$27.36                & 24.28$\pm$0.07  & 21.94$\pm$0.01  & 20.83$\pm$0.02  &  5.42  &  3.08  &    1.11  & 0.90   &  5.06     & QSO\\ 
10 51 19.41  & $-$05 55 25.7  & F4p32  & 24.17$\pm$0.05          & 22.74$\pm$0.02  & 21.94$\pm$0.01  & 22.14$\pm$0.04  &  2.23  &  1.43  & $-$0.20  & 0.93   & \nodata  & \nodata\\ 
10 51 54.74  & $-$05 46 26.8  & F4p32  & 24.87$\pm$0.09          & 22.97$\pm$0.02  & 22.01$\pm$0.01  & 21.70$\pm$0.03  &  2.86  &  1.90  &    0.31  & 1.03   & \nodata  & \nodata\\ 
10 52 19.78  & $-$05 28 18.2  & F4p32  & 25.35$\pm$0.14          & 23.66$\pm$0.03  & 22.92$\pm$0.01  & 22.92$\pm$0.08  &  2.43  &  1.70  &    0.00  & 0.93   &  3.89     & QSO\\ 
10 52 27.95  & $-$05 42 34.7  & F4p32  & 22.84$\pm$0.02          & 20.64$\pm$0.00  & 19.68$\pm$0.00  & 19.13$\pm$0.00  &  3.16  &  2.19  &    0.55  & 0.88   &  3.89     & QSO\\ 
10 53 48.89  & $-$05 33 19.4  & F4p31  & 23.82$\pm$0.03          & 21.65$\pm$0.01  & 20.94$\pm$0.00  & 20.54$\pm$0.01  &  2.88  &  2.17  &    0.40  & 0.99   & 4.20     & QSO\\ 
10 55 07.12  & $-$05 30 14.9  & F4p31  & 25.61$\pm$0.15          & 22.37$\pm$0.01  & 21.02$\pm$0.00  & 20.35$\pm$0.01  &  4.59  &  3.23  &    0.67  & 0.93   &  4.40     & QSO\\ 
10 55 23.03  & $-$05 48 50.7  & F4p31  & 26.27$\pm$0.27          & 23.34$\pm$0.03  & 22.35$\pm$0.01  & 21.82$\pm$0.03  &  3.92  &  2.92  &    0.52  & 0.90   &  4.12     & QSO\\ 
10 55 44.41  & $-$05 31 55.9  & F4p31  & 24.44$\pm$0.05          & 23.11$\pm$0.02  & 22.06$\pm$0.01  & 22.71$\pm$0.06  &  2.38  &  1.34  & $-$0.65  & 0.96   &  \nodata  & \nodata \\ 
\enddata



\end{deluxetable}


\begin{deluxetable}{cccccccccccc}
\tablecolumns{12}

\rotate

\tabletypesize{\footnotesize}

\tablewidth{0pt}

\tablecaption{$z\sim 4$ Quasar Sample \label{tab:qso_fsel} }


\tablehead{
\colhead{} & \colhead{} & \colhead{} & \colhead{} & \colhead{} & \multicolumn{3}{c}{Simulated, $z$, $I$} & \multicolumn{3}{c}{Spectrum} & \colhead{} \\
\colhead{Name} & \colhead{$R$} & \colhead{$z$} & \colhead{$f_{\rm sel}$} & \colhead{$f_{\rm spec}$} & \colhead{$M_{1450}$} & \colhead{$z_{\rm lim}$\tablenotemark{a}} & \colhead{$V_a$\tablenotemark{a}} & \colhead{$M_{1450}$} & \colhead{$z_{\rm lim}$\tablenotemark{a}} & \colhead{$V_a$\tablenotemark{a}} & \colhead{Area} \\ 
\colhead{} & \colhead{(mag)} & \colhead{} & \colhead{} & \colhead{} & \colhead{(mag)} & \colhead{} & \colhead{(Mpc$^3$ arcmin$^{-2}$)} & \colhead{(mag)} & \colhead{} & \colhead{(Mpc$^3$ arcmin$^{-2}$)} & \colhead{(deg$^2$)} \\
\colhead{(1)} & \colhead{(2)} & \colhead{(3)} & \colhead{(4)} & \colhead{(5)} & 
\colhead{(6)} & \colhead{(7)} & \colhead{(8)} & \colhead{(9)} & \colhead{(10)} & \colhead{(11)} & \colhead{(12)} } 

\startdata
\cutinhead{NDWFS}
NDWFS J142522.71+324827.0 &  21.10 &  3.81 &  0.723 &  0.75 & $-$24.74 & 3.81 & 4024.4 &$-$23.79 & 3.81 & 2588.5 & 1.71 \\
NDWFS J142713.19+322840.8 &  23.70 &  3.74 &  0.387 &  0.036 &$-$22.17 & 3.74\tablenotemark{c} & 1248.8\tablenotemark{c} &$-$21.46 & 3.74\tablenotemark{c} &  278.5\tablenotemark{c} & 1.71 \\
NDWFS J143017.78+322003.4 &  22.23 &  4.75 &  0.655 &  0.75 & $-$24.80 & 4.75 & 4104.7 &$-$23.89 & 4.75 & 2786.6 & 1.71 \\
NDWFS J143101.94+325155.5 &  21.77 &  3.94 &  0.806 &  0.75 & $-$24.60 & 3.94 & 3862.9 &$-$24.90 & 3.94 & 4238.0 & 1.71 \\
NDWFS J143324.54+340841.2 &  22.62 &  3.88 &  0.708 &  0.75 & $-$23.13 & 3.88 & 1159.2 &$-$22.63 & 3.88 &  278.5 & 1.71 \\
NDWFS J143331.15+343248.3 &  20.52 &  4.15 &  0.842 &  0.75 & $-$25.37 & 4.15 & 4291.1 &$-$25.54 & 4.15 & 4291.1 & 1.71 \\
NDWFS J143642.86+350923.8 &  22.00 &  3.90 &  0.782 &  0.75 & $-$23.80 & 3.90 & 2616.9 &$-$23.29 & 3.90 & 1516.1 & 1.71 \\
NDWFS J143658.34+333632.0 &  20.43 &  4.02 &  0.839 &  0.75 & $-$25.51 & 4.02 & 4291.1 &$-$25.26 & 4.02 & 4291.1 & 1.71 \\
NDWFS J143732.67+335522.0 &  22.88 &  4.22 &  0.724 &  0.75 & $-$23.36 & 4.22 & 1692.9 &$-$23.01 & 4.22 &  858.4 & 1.71 \\
NDWFS J143813.85+350236.4 &  22.85 &  4.25 &  0.726 &  0.75 & $-$23.28 & 4.25 & 1486.5 &$-$22.95 & 4.25 &  737.2 & 1.71 \\
NDWFS J143839.68+351245.9 &  21.10 &  4.63 &  0.805 &  0.75 & $-$25.65 & 4.63 & 4291.1 &$-$25.15 & 4.63 & 4291.1 & 1.71 \\
\cutinhead{DLS}
  DLS J091527.53+291750.4 &  20.72 &  4.34 &  0.909 &  0.714 &$-$25.84 & 4.34 & 4291.1 &$-$25.60 & 4.34 & 4291.1 & 2.05 \\
  DLS J092151.96+292457.2 &  22.46 &  4.32 &  0.805 &  0.714 &$-$24.13 & 4.32 & 3178.3 &$-$23.75 & 4.32 & 2531.6 & 2.05 \\
  DLS J092236.49+301010.6 &  22.90 &  3.98 &  0.497 &  0.714 &$-$22.86 & 3.98 &  462.8 &$-$22.11 & 3.98 &  278.5 & 2.05 \\
  DLS J092327.22+295251.7 &  21.58 &  3.84 &  0.090 &  0.714 &$-$24.05 & 3.84 & 3039.1 &$-$23.11 & 3.84 & 1129.2 & 2.05 \\
  DLS J092336.82+300949.9 &  21.94 &  5.06 &  0.739 &  0.714 &$-$25.54 & 5.06 & 4291.1 &$-$25.22 & 5.06 & 4291.1 & 2.05 \\
DLS J105011.52$-$044253.9 &  23.24 &  4.27 &  0.737 &  0.125 &$-$23.07 & 4.27\tablenotemark{b} & 2158.9\tablenotemark{b} &$-$22.70 & 4.27\tablenotemark{b} & 1338.2\tablenotemark{b} & 2.05 \\
DLS J105219.78$-$052818.2 &  22.92 &  3.89 &  0.228 &  0.714 &$-$22.82 & 3.89 &  340.0 &$-$22.51 & 3.89 &  278.5 & 2.05 \\
DLS J105227.95$-$054234.7 &  19.68 &  3.89 &  0.302 &  0.714 &$-$26.61 & 3.89 & 4291.1 &$-$25.52 & 3.89 & 4291.1 & 2.05 \\
DLS J105346.14$-$052859.5 &  23.83 &  4.02 &  0.371 &  0.021 &$-$20.40 & 4.02\tablenotemark{c} &  278.5\tablenotemark{c} &$-$21.74 & 4.02\tablenotemark{c} &  278.5\tablenotemark{c} & 2.05 \\
DLS J105348.89$-$053319.4 &  20.94 &  4.20 &  0.924 &  0.714 &$-$25.38 & 4.20 & 4291.1 &$-$24.78 & 4.20 & 4078.0 & 2.05 \\
DLS J105507.12$-$053014.9 &  21.02 &  4.40 &  0.891 &  0.714 &$-$25.68 & 4.40 & 4291.1 &$-$25.18 & 4.40 & 4291.1 & 2.05 \\
DLS J105523.03$-$054850.7 &  22.35 &  4.12 &  0.779 &  0.714 &$-$24.06 & 4.12 & 3067.0 &$-$24.40 & 4.12 & 3564.0 & 2.05 \\
\enddata

\tablenotetext{a}{These values are computed to a survey magnitude limit of $R=23$, unless otherwise specified.}
\tablenotetext{b}{These values were computed to a survey magnitude limit of $R=23.5$.}
\tablenotetext{c}{These values were computed to a survey magnitude limit of $R=24$.}



\end{deluxetable}


\begin{deluxetable}{cccccccc}
\tablecolumns{8}

\rotate

\tabletypesize{\footnotesize}

\tablewidth{0pt}
\tablecaption{Luminosity Function\label{tab:lumfunc}}


\tablehead{\colhead{} & \multicolumn{2}{c}{NDWFS} & \multicolumn{2}{c}{DLS} & \colhead{} &\multicolumn{2}{c}{All Quasars} \\
\colhead{$M_{1450}$ bin center} & \colhead{$\Phi$} & \colhead{$N_{\rm QSO}$} & \colhead{$\Phi$} & \colhead{$N_{\rm QSO}$} & \colhead{$<M_{1450}>$ } & \colhead{$\Phi$} & \colhead{$N_{QSO}$} \\ 
\colhead{(mag)} & \colhead{(10$^{-7}$ Mpc$^{-3}$ mag$^{-1}$)} & \colhead{} & \colhead{(10$^{-7}$ Mpc$^{-3}$ mag$^{-1}$)} & \colhead{} & \colhead{(mag)} & \colhead{(10$^{-7}$ Mpc$^{-3}$ mag$^{-1}$)} & \colhead{} } 

\startdata
\cutinhead{$M_{1450}$ from simulated quasars and $z$, $I$ magnitudes}
$-$26.5 & \nodata & \nodata & 1.46$\pm$1.46 & 1 & $-$26.61 & 0.80$\pm$0.80 & 1 \\
$-$25.5 & 1.83$\pm$1.06 & 3 & 2.06$\pm$1.03 & 4 & $-$25.58 & 1.95$\pm$0.74 & 7 \\
$-$24.5 & 2.25$\pm$1.30 & 3 & 8.48$\pm$7.03 & 3 & $-$24.44 & 5.65$\pm$3.88 & 6 \\
$-$23.5 & 7.47$\pm$3.90 & 4 & 6.81$\pm$6.81 & 1 & $-$23.36 & 7.11$\pm$4.12 & 5 \\
$-$22.5 & 92.4$\pm$92.4 & 1 & 32.7$\pm$25.8 & 2 & $-$22.66 & 59.9$\pm$44.3 & 3 \\
$-$21.5 & \nodata & \nodata & \nodata & \nodata & \nodata & \nodata & \nodata \\
$-$20.5 & \nodata & \nodata & 630$\pm$630 & 1 & $-$20.40 & 344$\pm$344 & 1 \\
\cutinhead{$M_{1450}$ based on object spectra}
$-$25.5 & 1.83$\pm$1.06 & 3 & 3.05$\pm$1.73 & 4 & $-$25.37 & 2.49$\pm$1.06 & 7 \\
$-$24.5 & 0.63$\pm$0.63 & 1 & 1.19$\pm$0.85 & 2 & $-$24.71 & 0.93$\pm$0.55 & 3 \\
$-$23.5 & 7.66$\pm$4.27 & 4 & 19.6$\pm$18.7 & 2 & $-$23.53 & 14.2$\pm$10.4 & 6 \\
$-$22.5 & 15.0$\pm$11.7 & 2 & 54.6$\pm$34.7 & 3 & $-$22.61 & 36.6$\pm$19.6 & 5 \\
$-$21.5 & 414$\pm$414 & 1 & 630$\pm$630 & 1 & $-$21.61 & 532$\pm$392 & 2 \\
\enddata




\end{deluxetable}


\begin{deluxetable}{cc}



\tablewidth{0pt}
\tablecaption{Single Power Law Fit \label{tab:splparam}}


\tablehead{\colhead{Dataset} & \colhead{$\beta$} }

\startdata
$z$,$I$, $R<23$   &  $-1.59\pm0.22$ \\ 
spec, $R<23$      &  $-1.23\pm0.20$ \\ 
$z$,$I$, all QSOs &  $-1.98\pm0.23$ \\ 
spec, all QSOs    &  $-2.46\pm0.20$ \\ 
\enddata




\end{deluxetable}


\begin{deluxetable}{cccc}



\tablewidth{0pt}
\tablecaption{Double-Power-Law Parameters \label{tab:dplparam}}


\tablehead{\colhead{} & \colhead{Fontanot+$z$,$I$} & \colhead{Fontanot+spec} & \colhead{Richards+spec} } 

\startdata
Dataset & (1) & (2) & (3)\\
$\Phi^*$\tablenotemark{a} & 2.5$\pm$5.8$\times10^{-8}$ & 7.1$\pm$2.9$\times10^{-8}$ & 5.6$\pm$1.0$\times10^{-7}$ \\
$\alpha$                                & $-$3.1$\pm$0.1                        & $-$3.6$\pm$0.3                       & $-$2.4$\pm$0.2 \\
$\beta$                                  & $-$1.4$\pm$0.1                        & $-$1.6$\pm$0.2                       & $-$2.3$\pm$0.2 \\
$M^*$\tablenotemark{b}    & $-$25.6$\pm$0.2                      & $-$26.6$\pm$0.3                    & $-$24.1$\pm$0.1 \\
$\chi^2_{\rm reduced}$      & 0.73                                            & 0.90                                            & 1.18 \\
\enddata
\tablenotetext{a}{Mpc$^{-3}$ mag$^{-1}$}
\tablenotetext{b}{mag$^{-1}$}


\end{deluxetable}

\begin{figure}
\plotone{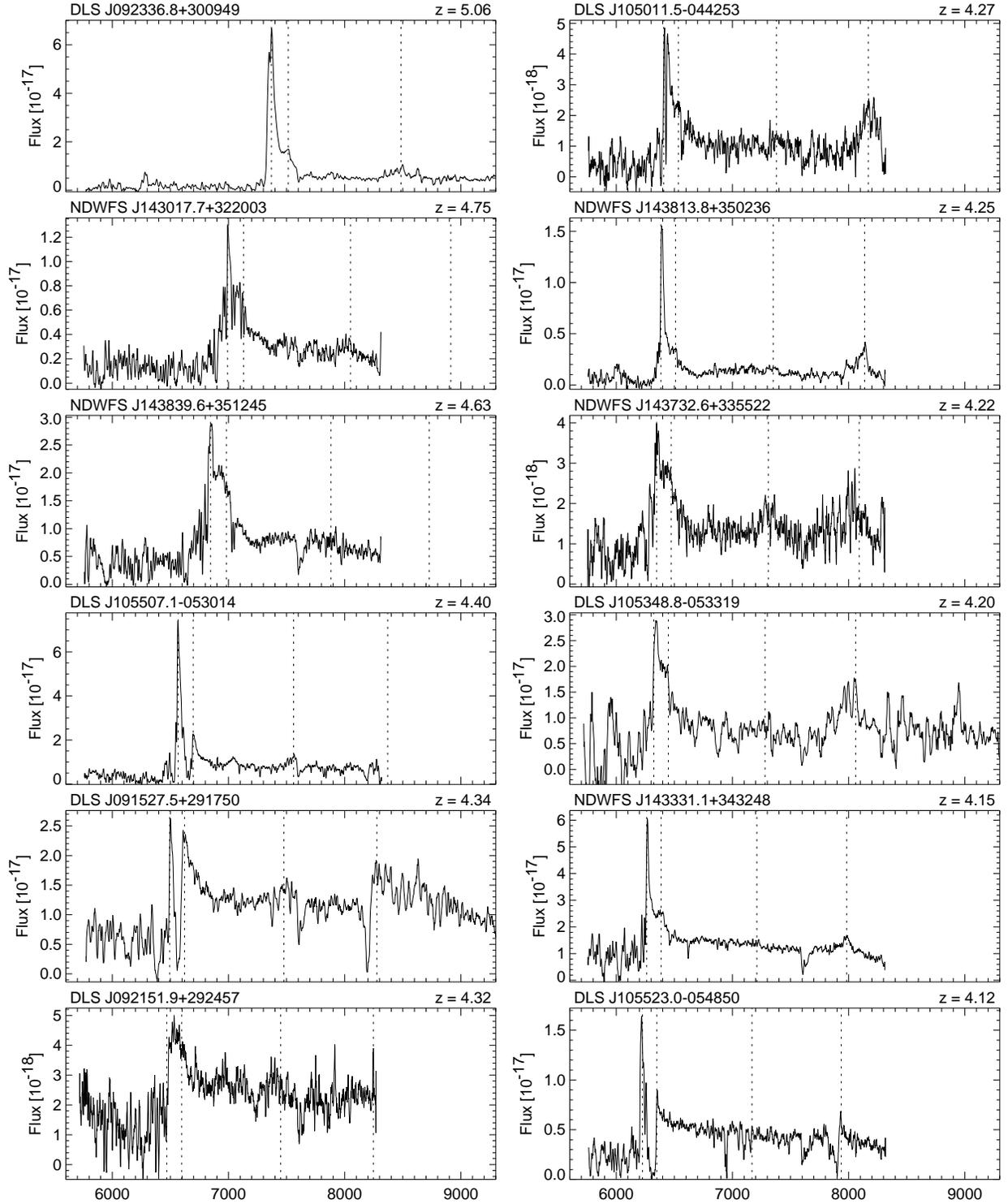}
\caption{
Spectra of DLS and NDWFS quasars, ordered by
redshift.  The dotted lines show expected positions of
prominent emission lines in the ultraviolet:
Ly$\alpha$~1216,
N~V~1240,
Si~IV~1400,
C~IV~1550,
C~III]~1909.
}\label{fig:spectra}
\end{figure}

\begin{figure}
\plotone{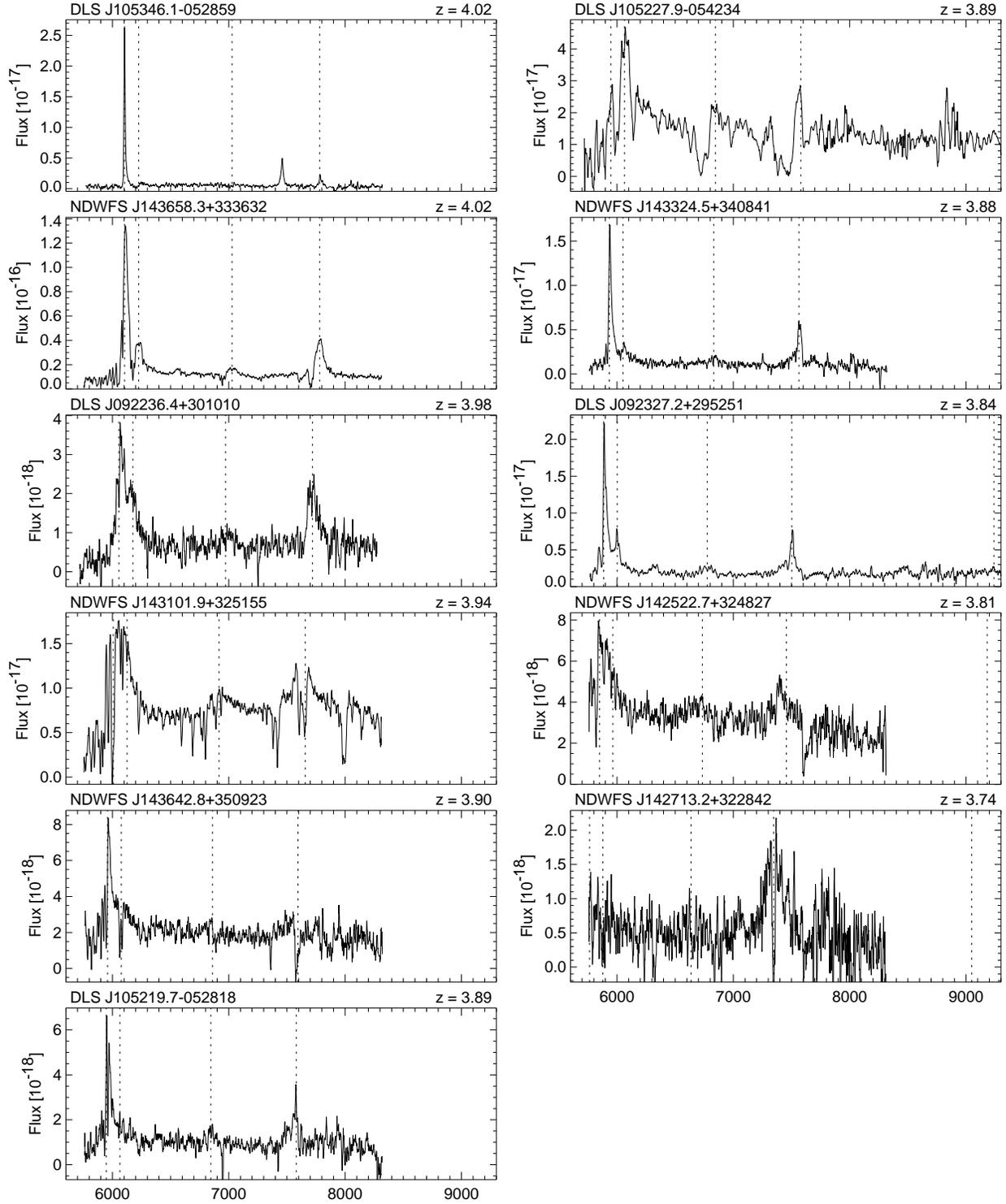}
\caption{{\it Continued.} Spectra of DLS and NDWFS quasars.}\label{fig:spectra2}
\end{figure}

 \begin{figure}
   \plotone{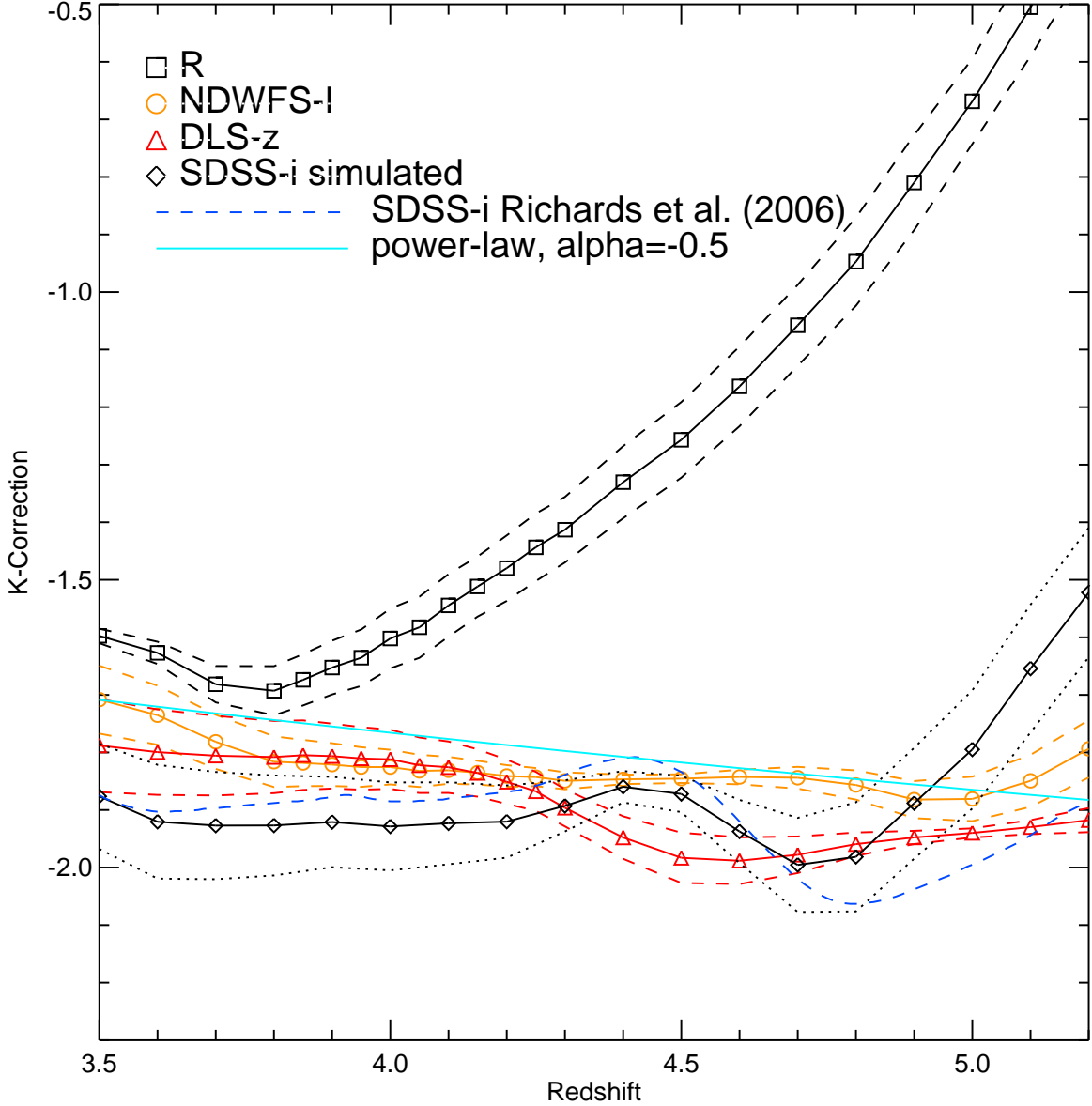}
   \caption{K-correction determined using simulated quasar spectra. The black squares present the value of the correction for the NDWFS and DLS $R$-band, while the orange circles are the correction for the NDWFS I-band and the red triangles are the corrections for the DLS z-band.  The $\pm1\sigma$ spread is plotted with the dashed curve in the corresponding color. The cyan solid line is the K-correction one would obtain using a fixed value of a power-law slope $\alpha=-0.5$ and no emission line or {\lya} forest contribution (equation \ref{kcorr-nu}). The blue dashed curve is the K-correction of \citet{Richards06} for the SDSS $i'$-band. The black diamonds and curve present the K-correction for the SDSS $i'$-band determined using our simulated quasar spectra, with the $\pm1\sigma$ intervals outlined by the red dash-dot curve. As Ly$\alpha$ moves through the bandpass, the difference between the correction to the R-band and the redder bands increases significantly.}
   \label{fig:kcorr}
 \end{figure}

 \begin{figure}
   \plotone{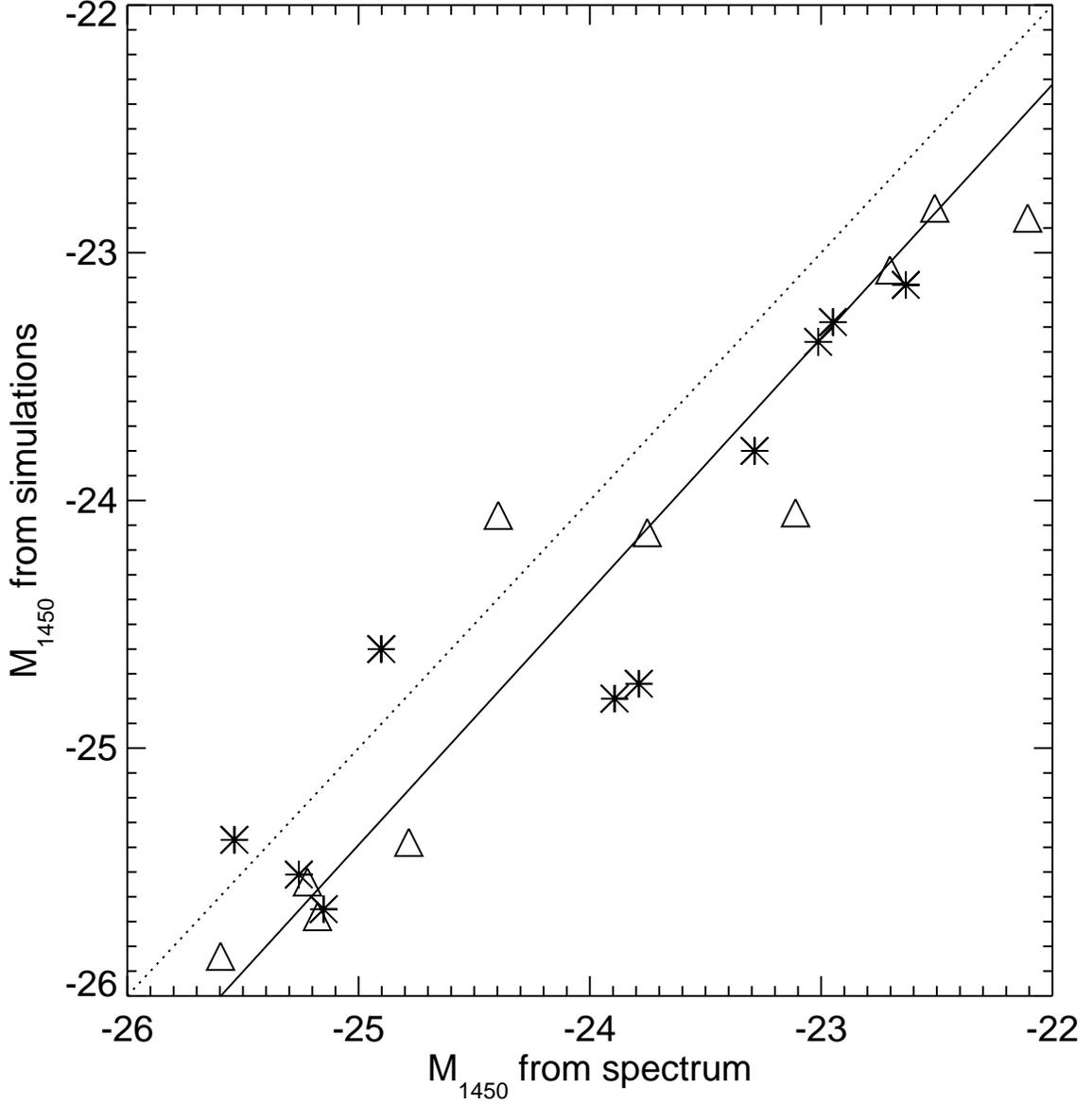}
   \caption{Absolute magnitude at rest-frame 1450\AA, $M_{1450}$, computed using the K-correction derived from our library of simulated quasar spectra versus $M_{1450}$ measured directly from our spectra. DLS quasars are marked with {\em asterisks} and NDWFS quasars are marked with {\em triangles}. The solid line is the best fit relationship, with a slope of $1.02\pm0.09$ and an offset of $0.19\pm2.06$, suggesting that $M_{1450}$ measured from the spectra are systematically fainter by $\sim 0.3$ magnitudes. The dotted line represents a one-to-one relationship.}
   \label{fig:absmag}
 \end{figure}

\begin{figure}
\plottwo{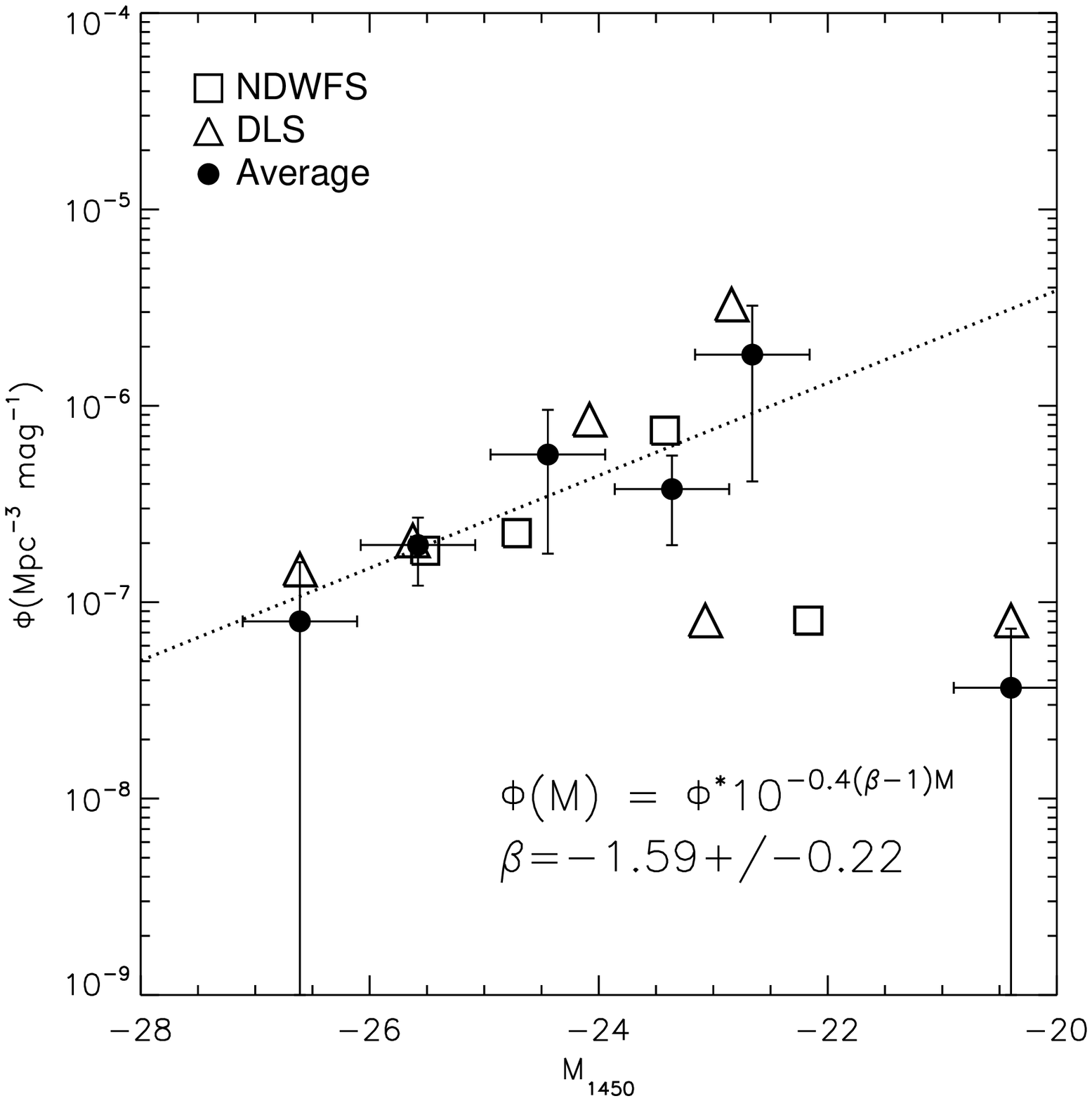}{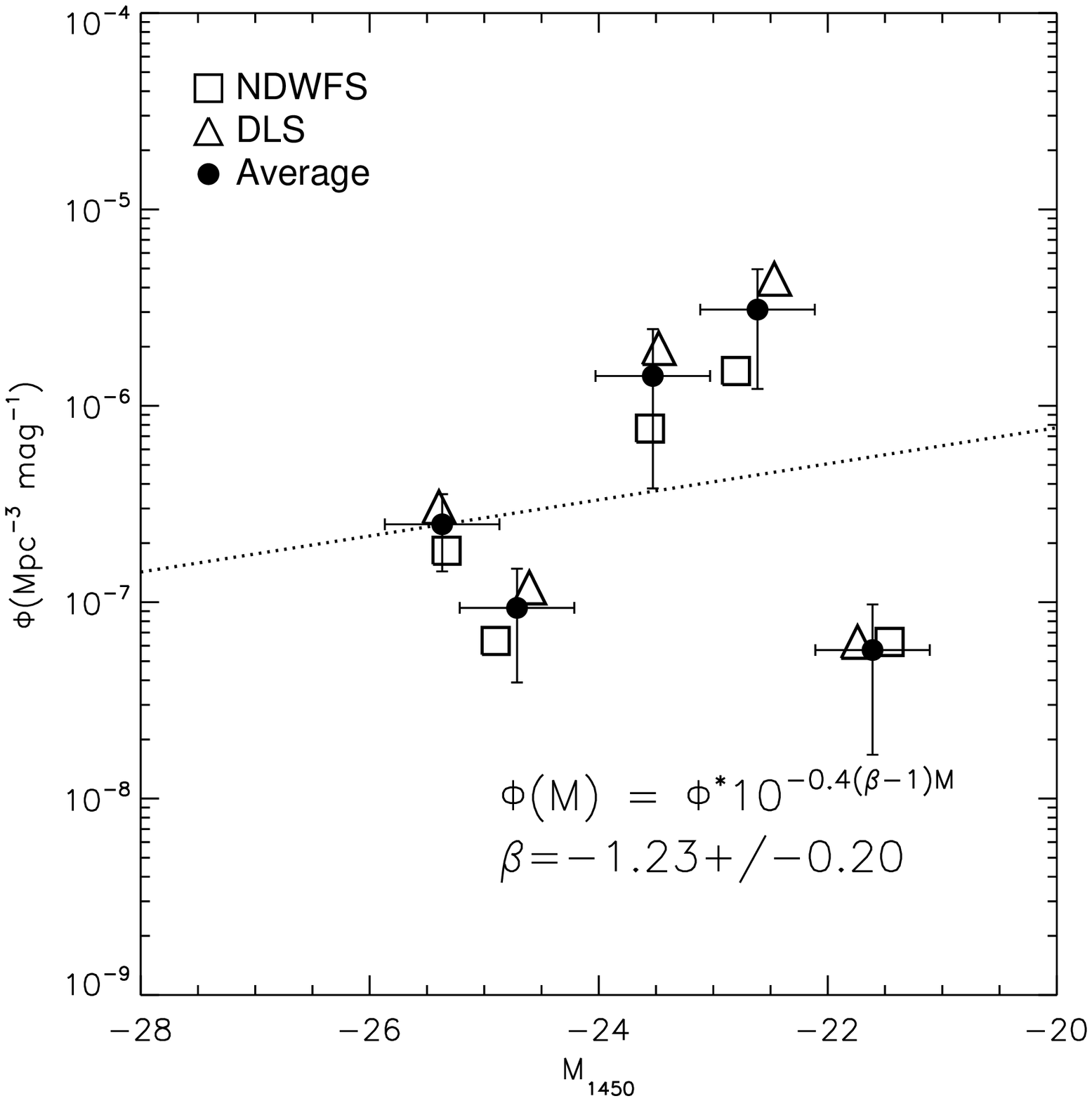}
\caption{The comoving volume density computed for the quasars in our survey with $R \leq 23$.   Open squares are $\Phi(M_{1450},z=4.15)$ computed from quasars in the NDWFS survey, open triangles are for quasars in the DLS survey, and the filled circles and their corresponding error bars are the comoving volume density of the quasars from both surveys combined. The left-hand panel is the QLF based on $M_{1450}$ computed from the $z$ (for DLS) and $I$ (for NDWFS) bands.  The right-hand panel is the QLF computed using the individual quasar spectra to compute $M_{1450}$ (as described in \S \ref{sec:kcorr}). The dotted line is a single power-law fit, $\Phi(M) = \Phi^* 10^{-0.4(\beta-1)M}$, to the measurements of the QLF using the combined survey values.  }\label{fig:lf1}
\end{figure}

\begin{figure}
\plottwo{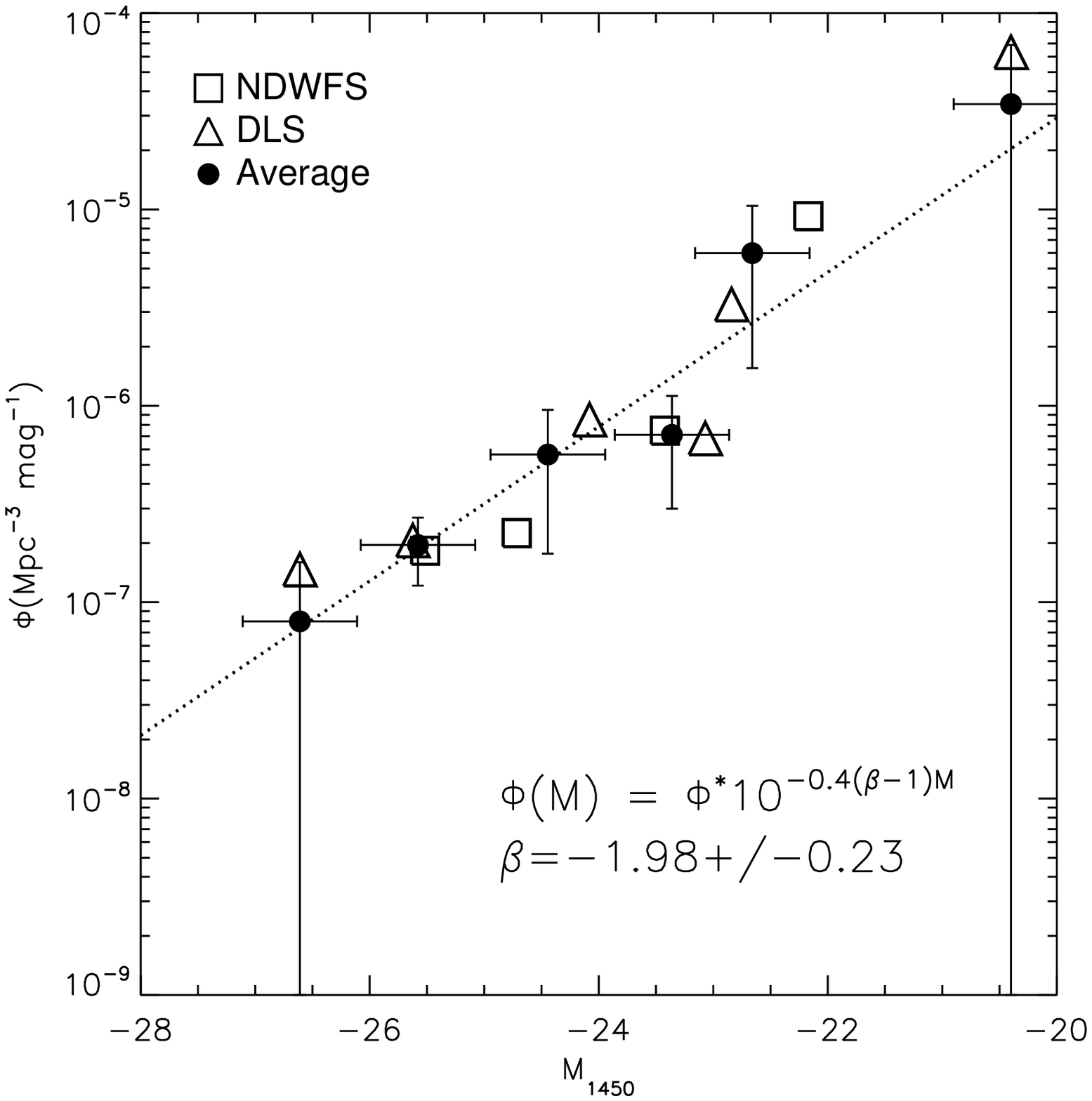}{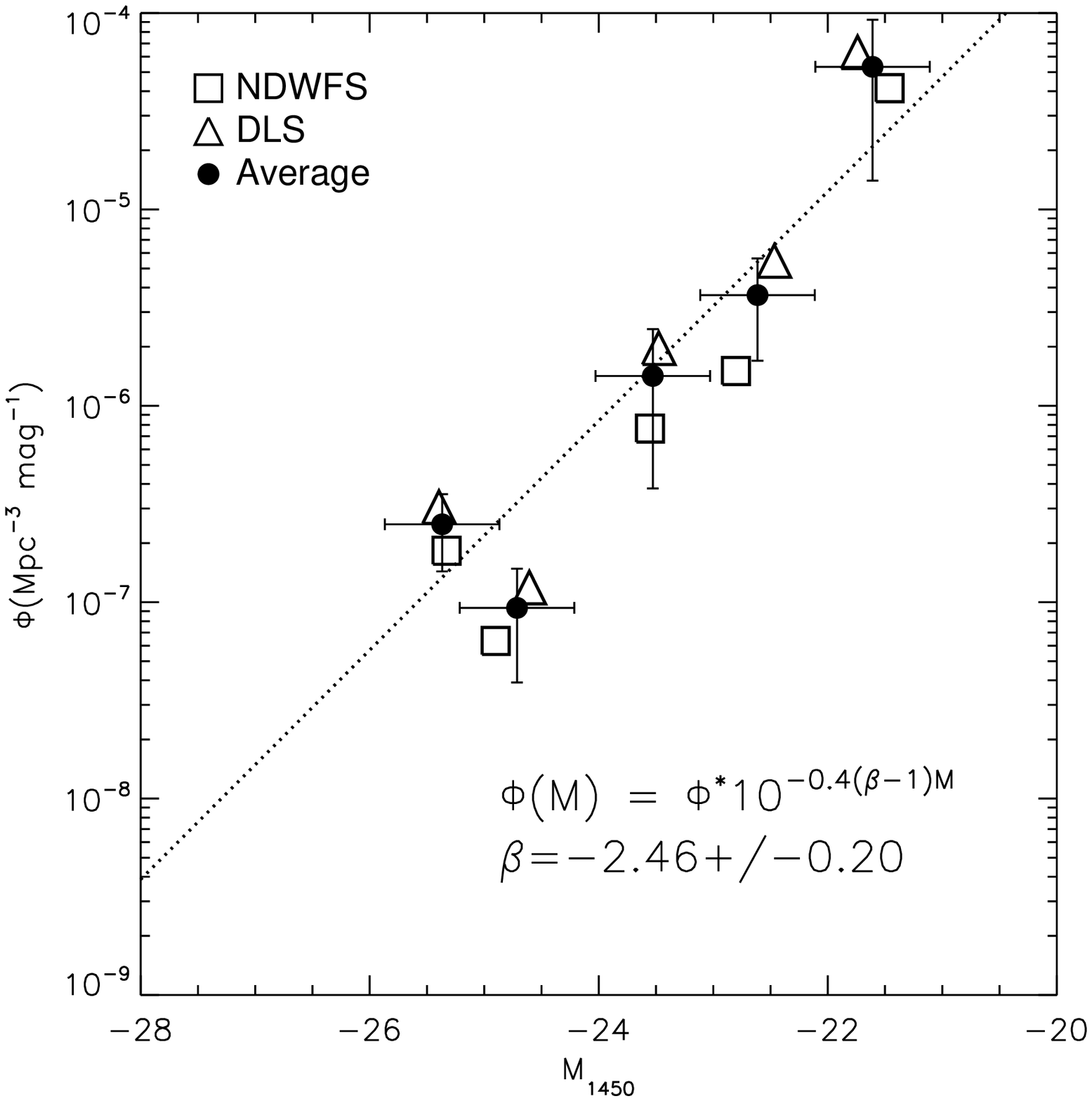}
\caption{The comoving volume density computed for all the quasars in our survey.   The symbols and lines are the same as in Figure \ref{fig:lf1} as are the descriptions of the left and right panels.   }\label{fig:lf2}
\end{figure}

\begin{figure}
\plottwo{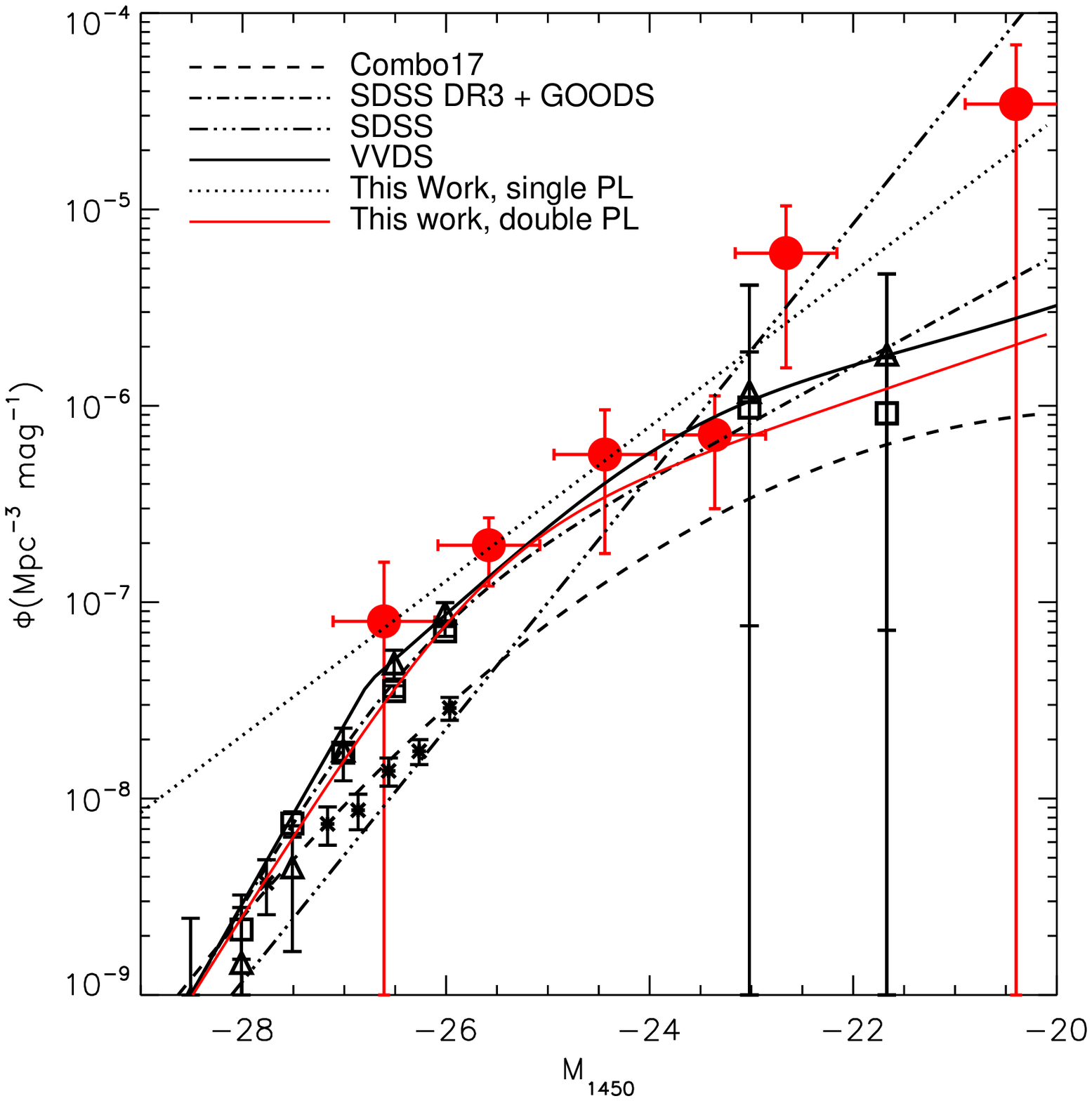}{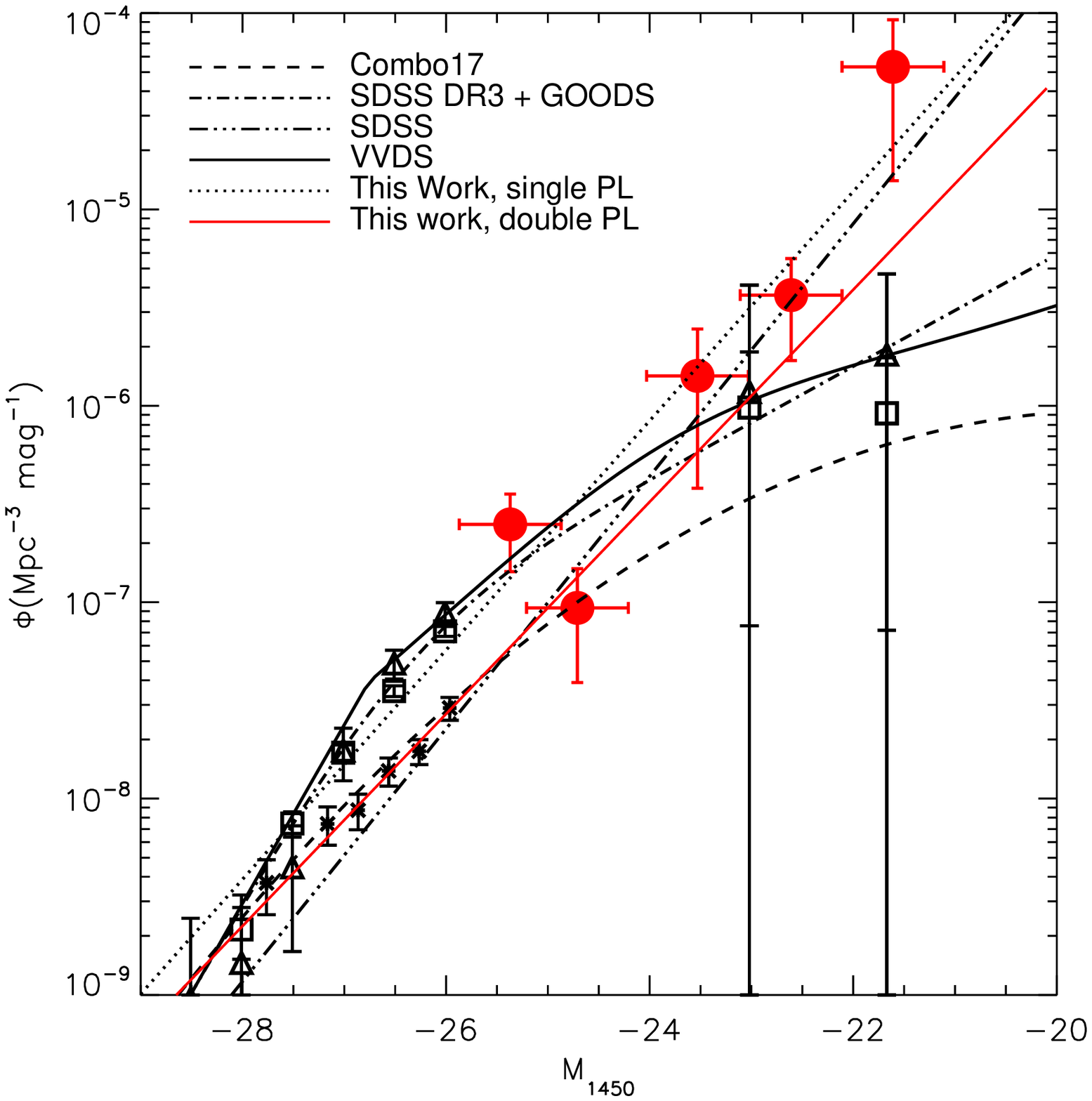}
\caption{The $z=4.15$ quasar luminosity function.  Asterisks are the $z=4.25$ binned data from \citet{Richards06}, the open triangles and squares are the combined SDSS and GOODS QLF from \citet{Fontanot07a} using at $z=3.75$ and $z=4.6$, respectively.  Our binned QLF data are plotted with red circles, where $M_{1450}$ has been computed from K-corrections to the $z$ and $I$ band photometry in the left-hand panel and directly from the spectra in the right-hand panel.  The filled circles are the mean NDWFS and DLS measurements, while the open circle is the faint bin made up of two quasars from the DLS survey.  We plot the functional form of the QLF from various sources (summarized in the legend).  All the points and curves have been evolved to $z=4.15$; see text for details.  }\label{fig:qlf_all}
\end{figure}

\begin{figure}
\plotone{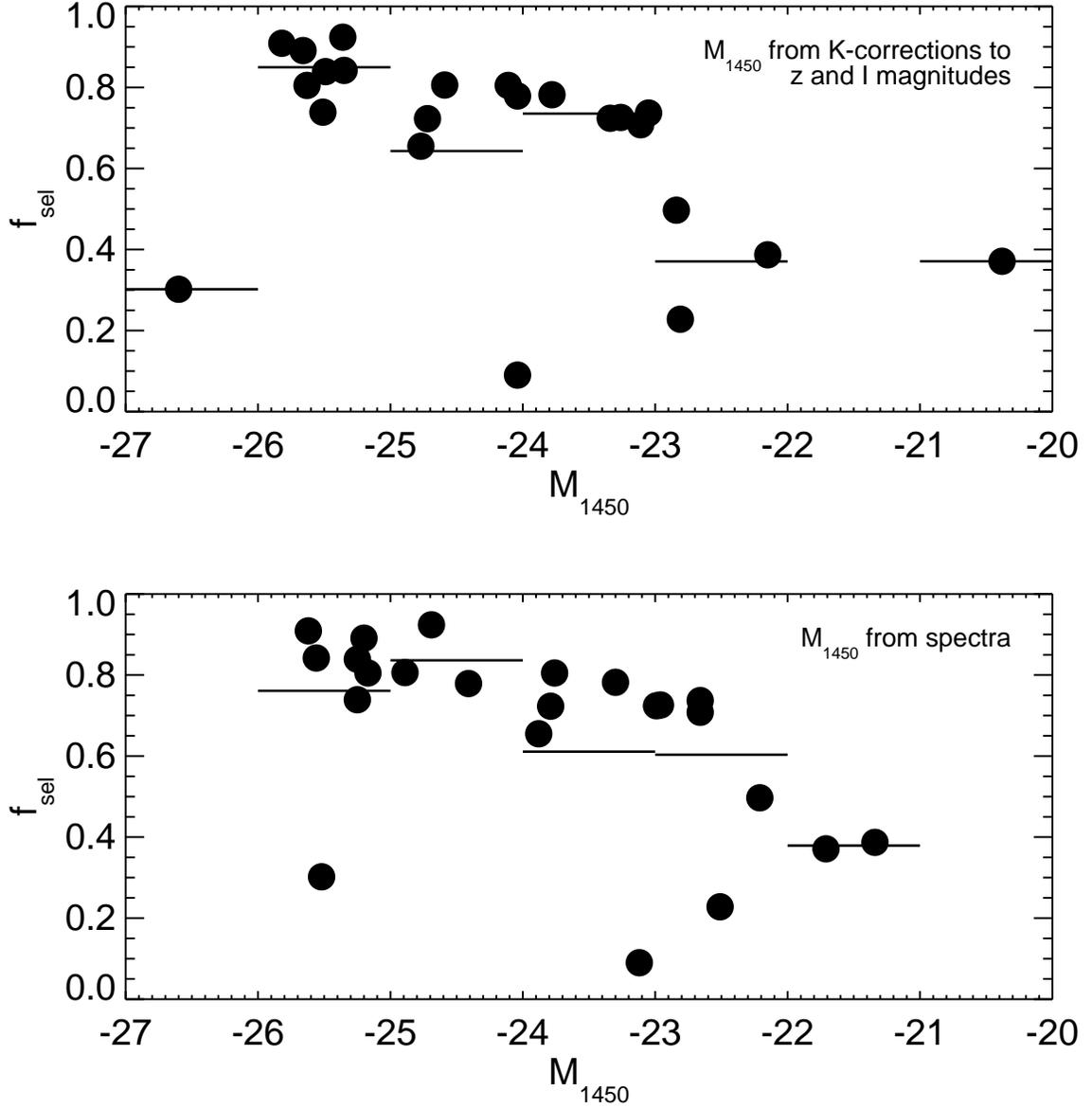}
\caption{Our selection completeness, $f_{\rm sel}$, as a function of the quasar's absolute magnitude at 1450\AA, $M_{1450}$.  The top panel plots the objects with $M_{1450}$ derived from the K-corrections to $z$ and $I$ magnitudes and the bottom panel plots the objects with $M_{1450}$ measured from their spectra.  The horizontal lines are the mean value of the selection function for each luminosity bin.}\label{fig:fselvm1450}
\end{figure}


\begin{thebibliography}{55}
\expandafter\ifx\csname natexlab\endcsname\relax\def\natexlab#1{#1}\fi

\bibitem[{{Baldwin} {et~al.}(2003){Baldwin}, {Hamann}, {Korista}, {Ferland},
  {Dietrich}, \& {Warner}}]{Baldwin03}
{Baldwin}, J.~A., {Hamann}, F., {Korista}, K.~T., {Ferland}, G.~J., {Dietrich},
  M., \& {Warner}, C. 2003, \apj, 583, 649

\bibitem[{{Barden} {et~al.}(2000){Barden}, {Harmer}, {Blakley}, \&
  {Parks}}]{Barden01}
{Barden}, S.~C., {Harmer}, C.~F., {Blakley}, R.~D., \& {Parks}, R.~J. 2000, in
  Society of Photo-Optical Instrumentation Engineers (SPIE) Conference Series,
  Vol. 4008, Society of Photo-Optical Instrumentation Engineers (SPIE)
  Conference Series, ed. {M.~Iye \& A.~F.~Moorwood}, 593--600

\bibitem[{{Bentz} {et~al.}(2004){Bentz}, {Hall}, \& {Osmer}}]{Bentz04}
{Bentz}, M.~C., {Hall}, P.~B., \& {Osmer}, P.~S. 2004, \aj, 128, 561

\bibitem[{{Bershady} {et~al.}(1999){Bershady}, {Charlton}, \&
  {Geoffroy}}]{Bershady99}
{Bershady}, M.~A., {Charlton}, J.~C., \& {Geoffroy}, J.~M. 1999, \apj, 518, 103

\bibitem[{{Bertin} \& {Arnouts}(1996)}]{Bertin96}
{Bertin}, E., \& {Arnouts}, S. 1996, \aaps, 117, 393

\bibitem[{{Blanton} {et~al.}(2003){Blanton}, {Brinkmann}, {Csabai}, {Doi},
  {Eisenstein}, {Fukugita}, {Gunn}, {Hogg}, \& {Schlegel}}]{Blanton03}
{Blanton}, M.~R. {et~al.} 2003, \aj, 125, 2348

\bibitem[{{Bongiorno} {et~al.}(2007){Bongiorno}, {Zamorani}, {Gavignaud},
  {Marano}, {Paltani}, {Mathez}, {M{\o}ller}, {Picat}, {Cirasuolo},
  {Lamareille}, {Bottini}, {Garilli}, {Le Brun}, {Le F{\`e}vre}, {Maccagni},
  {Scaramella}, {Scodeggio}, {Tresse}, {Vettolani}, {Zanichelli}, {Adami},
  {Arnouts}, {Bardelli}, {Bolzonella}, {Cappi}, {Charlot}, {Ciliegi},
  {Contini}, {Foucaud}, {Franzetti}, {Guzzo}, {Ilbert}, {Iovino}, {McCracken},
  {Marinoni}, {Mazure}, {Meneux}, {Merighi}, {Pell{\`o}}, {Pollo}, {Pozzetti},
  {Radovich}, {Zucca}, {Hatziminaoglou}, {Polletta}, {Bondi}, {Brinchmann},
  {Cucciati}, {de La Torre}, {Gregorini}, {Mellier}, {Merluzzi}, {Temporin},
  {Vergani}, \& {Walcher}}]{Bongiorno07}
{Bongiorno}, A. {et~al.} 2007, \aap, 472, 443

\bibitem[{{Boyle} {et~al.}(1988){Boyle}, {Shanks}, \& {Peterson}}]{Boyle88}
{Boyle}, B.~J., {Shanks}, T., \& {Peterson}, B.~A. 1988, \mnras, 235, 935

\bibitem[{{Cool} {et~al.}(2006){Cool}, {Kochanek}, {Eisenstein}, {Stern},
  {Brand}, {Brown}, {Dey}, {Eisenhardt}, {Fan}, {Gonzalez}, {Green}, {Jannuzi},
  {McKenzie}, {Rieke}, {Rieke}, {Soifer}, {Spinrad}, \& {Elston}}]{Cool06}
{Cool}, R.~J. {et~al.} 2006, \aj, 132, 823

\bibitem[{{Cristiani} \& {Vio}(1990)}]{Cristiani90}
{Cristiani}, S., \& {Vio}, R. 1990, \aap, 227, 385

\bibitem[{{Djorgovski}(2005)}]{Djorgovski05}
{Djorgovski}, S.~G. 2005, in The Tenth Marcel Grossmann Meeting., ed.
  M.~{Novello}, S.~{Perez Bergliaffa}, \& R.~{Ruffini}, 422--442

\bibitem[{{Elvis} {et~al.}(1994){Elvis}, {Wilkes}, {McDowell}, {Green},
  {Bechtold}, {Willner}, {Oey}, {Polomski}, \& {Cutri}}]{Elvis94}
{Elvis}, M. {et~al.} 1994, \apjs, 95, 1

\bibitem[{{Fan}(1999)}]{Fan99}
{Fan}, X. 1999, \aj, 117, 2528

\bibitem[{{Fan} {et~al.}(2001{\natexlab{a}}){Fan}, {Narayanan}, {Lupton},
  {Strauss}, {Knapp}, {Becker}, {White}, {Pentericci}, {Leggett}, {Haiman},
  {Gunn}, {Ivezi{\'c}}, {Schneider}, {Anderson}, {Brinkmann}, {Bahcall},
  {Connolly}, {Csabai}, {Doi}, {Fukugita}, {Geballe}, {Grebel}, {Harbeck},
  {Hennessy}, {Lamb}, {Miknaitis}, {Munn}, {Nichol}, {Okamura}, {Pier},
  {Prada}, {Richards}, {Szalay}, \& {York}}]{Fan01b}
{Fan}, X. {et~al.} 2001{\natexlab{a}}, \aj, 122, 2833

\bibitem[{{Fan} {et~al.}(2001{\natexlab{b}}){Fan}, {Strauss}, {Schneider},
  {Gunn}, {Lupton}, {Becker}, {Davis}, {Newman}, {Richards}, {White},
  {Anderson}, {Annis}, {Bahcall}, {Brunner}, {Csabai}, {Hennessy}, {Hindsley},
  {Fukugita}, {Kunszt}, {Ivezi{\'c}}, {Knapp}, {McKay}, {Munn}, {Pier},
  {Szalay}, \& {York}}]{Fan01a}
---. 2001{\natexlab{b}}, \aj, 121, 54

\bibitem[{{Ferrarese} \& {Merritt}(2000)}]{Ferrarese00}
{Ferrarese}, L., \& {Merritt}, D. 2000, \apjl, 539, L9

\bibitem[{{Fontanot} {et~al.}(2007){Fontanot}, {Cristiani}, {Monaco}, {Nonino},
  {Vanzella}, {Brandt}, {Grazian}, \& {Mao}}]{Fontanot07a}
{Fontanot}, F., {Cristiani}, S., {Monaco}, P., {Nonino}, M., {Vanzella}, E.,
  {Brandt}, W.~N., {Grazian}, A., \& {Mao}, J. 2007, \aap, 461, 39

\bibitem[{{Fosbury} {et~al.}(2003){Fosbury}, {Villar-Mart{\'{\i}}n},
  {Humphrey}, {Lombardi}, {Rosati}, {Stern}, {Hook}, {Holden}, {Stanford},
  {Squires}, {Rauch}, \& {Sargent}}]{Fosbury03}
{Fosbury}, R.~A.~E. {et~al.} 2003, \apj, 596, 797

\bibitem[{{Glikman} {et~al.}(2007){Glikman}, {Djorgovski}, {Stern},
  {Bogosavljevi{\'c}}, \& {Mahabal}}]{Glikman07a}
{Glikman}, E., {Djorgovski}, S.~G., {Stern}, D., {Bogosavljevi{\'c}}, M., \&
  {Mahabal}, A. 2007, \apjl, 663, L73

\bibitem[{{Haiman} {et~al.}(2001){Haiman}, {Abel}, \& {Madau}}]{Haiman01}
{Haiman}, Z., {Abel}, T., \& {Madau}, P. 2001, \apj, 551, 599

\bibitem[{{Ho}(2004)}]{Ho04}
{Ho}, L.~C., ed. 2004, {Coevolution of Black Holes and Galaxies}

\bibitem[{{Hopkins} {et~al.}(2005){Hopkins}, {Hernquist}, {Cox}, {Di Matteo},
  {Robertson}, \& {Springel}}]{Hopkins05a}
{Hopkins}, P.~F., {Hernquist}, L., {Cox}, T.~J., {Di Matteo}, T., {Robertson},
  B., \& {Springel}, V. 2005, \apj, 630, 716

\bibitem[{{Hopkins} {et~al.}(2006{\natexlab{a}}){Hopkins}, {Hernquist}, {Cox},
  {Robertson}, {Di Matteo}, \& {Springel}}]{Hopkins06b}
{Hopkins}, P.~F., {Hernquist}, L., {Cox}, T.~J., {Robertson}, B., {Di Matteo},
  T., \& {Springel}, V. 2006{\natexlab{a}}, \apj, 639, 700

\bibitem[{{Hopkins} {et~al.}(2006{\natexlab{b}}){Hopkins}, {Somerville},
  {Hernquist}, {Cox}, {Robertson}, \& {Li}}]{Hopkins06c}
{Hopkins}, P.~F., {Somerville}, R.~S., {Hernquist}, L., {Cox}, T.~J.,
  {Robertson}, B., \& {Li}, Y. 2006{\natexlab{b}}, \apj, 652, 864

\bibitem[{{Hunt} {et~al.}(2004){Hunt}, {Steidel}, {Adelberger}, \&
  {Shapley}}]{Hunt04}
{Hunt}, M.~P., {Steidel}, C.~C., {Adelberger}, K.~L., \& {Shapley}, A.~E. 2004,
  \apj, 605, 625

\bibitem[{{Ivezi{\'c}} {et~al.}(2004){Ivezi{\'c}}, {Lupton}, {Schlegel},
  {Boroski}, {Adelman-McCarthy}, {Yanny}, {Kent}, {Stoughton}, {Finkbeiner},
  {Padmanabhan}, {Rockosi}, {Gunn}, {Knapp}, {Strauss}, {Richards},
  {Eisenstein}, {Nicinski}, {Kleinman}, {Krzesinski}, {Newman}, {Snedden},
  {Thakar}, {Szalay}, {Munn}, {Smith}, {Tucker}, \& {Lee}}]{Ivezic04}
{Ivezi{\'c}}, {\v Z}. {et~al.} 2004, Astronomische Nachrichten, 325, 583

\bibitem[{{Jannuzi} \& {Dey}(1999)}]{Jannuzi99}
{Jannuzi}, B.~T., \& {Dey}, A. 1999, in ASP Conf. Ser. 191: Photometric
  Redshifts and the Detection of High Redshift Galaxies, ed. R.~{Weymann}
  {et~al.}, 111

\bibitem[{{Jiang} {et~al.}(2008){Jiang}, {Fan}, \& {Vestergaard}}]{Jiang08}
{Jiang}, L., {Fan}, X., \& {Vestergaard}, M. 2008, \apj, 679, 962

\bibitem[{{Kennefick} {et~al.}(1995{\natexlab{a}}){Kennefick}, {de Carvalho},
  {Djorgovski}, {Wilber}, {Dickson}, {Weir}, {Fayyad}, \&
  {Roden}}]{Kennefick95a}
{Kennefick}, J.~D., {de Carvalho}, R.~R., {Djorgovski}, S.~G., {Wilber}, M.~M.,
  {Dickson}, E.~S., {Weir}, N., {Fayyad}, U., \& {Roden}, J.
  1995{\natexlab{a}}, \aj, 110, 78

\bibitem[{{Kennefick} {et~al.}(1995{\natexlab{b}}){Kennefick}, {Djorgovski}, \&
  {de Carvalho}}]{Kennefick95b}
{Kennefick}, J.~D., {Djorgovski}, S.~G., \& {de Carvalho}, R.~R.
  1995{\natexlab{b}}, \aj, 110, 2553

\bibitem[{{Madau} {et~al.}(1999){Madau}, {Haardt}, \& {Rees}}]{Madau99}
{Madau}, P., {Haardt}, F., \& {Rees}, M.~J. 1999, \apj, 514, 648

\bibitem[{{Magorrian} {et~al.}(1998){Magorrian}, {Tremaine}, {Richstone},
  {Bender}, {Bower}, {Dressler}, {Faber}, {Gebhardt}, {Green}, {Grillmair},
  {Kormendy}, \& {Lauer}}]{Magorrian98}
{Magorrian}, J. {et~al.} 1998, \aj, 115, 2285

\bibitem[{{Marchesini} {et~al.}(2007){Marchesini}, {van Dokkum}, {Quadri},
  {Rudnick}, {Franx}, {Lira}, {Wuyts}, {Gawiser}, {Christlein}, \&
  {Toft}}]{Marchesini07}
{Marchesini}, D. {et~al.} 2007, \apj, 656, 42

\bibitem[{{Massey} \& {Gronwall}(1990)}]{Massey90}
{Massey}, P., \& {Gronwall}, C. 1990, \apj, 358, 344

\bibitem[{{McGreer} {et~al.}(2006){McGreer}, {Becker}, {Helfand}, \&
  {White}}]{McGreer06}
{McGreer}, I.~D., {Becker}, R.~H., {Helfand}, D.~J., \& {White}, R.~L. 2006,
  \apj, 652, 157

\bibitem[{{Oke} {et~al.}(1995){Oke}, {Cohen}, {Carr}, {Cromer}, {Dingizian},
  {Harris}, {Labrecque}, {Lucinio}, {Schaal}, {Epps}, \& {Miller}}]{Oke95}
{Oke}, J.~B. {et~al.} 1995, \pasp, 107, 375

\bibitem[{{Osmer} \& {Smith}(1980)}]{Osmer80}
{Osmer}, P.~S., \& {Smith}, M.~G. 1980, \apjs, 42, 333

\bibitem[{{Pei}(1995)}]{Pei95}
{Pei}, Y.~C. 1995, \apj, 438, 623

\bibitem[{{Richards} {et~al.}(2002){Richards}, {Fan}, {Newberg}, {Strauss},
  {Vanden Berk}, {Schneider}, {Yanny}, {Boucher}, {Burles}, {Frieman}, {Gunn},
  {Hall}, {Ivezi{\'c}}, {Kent}, {Loveday}, {Lupton}, {Rockosi}, {Schlegel},
  {Stoughton}, {SubbaRao}, \& {York}}]{Richards02}
{Richards}, G.~T. {et~al.} 2002, \aj, 123, 2945

\bibitem[{{Richards} {et~al.}(2003){Richards}, {Hall}, {Vanden Berk},
  {Strauss}, {Schneider}, {Weinstein}, {Reichard}, {York}, {Knapp}, {Fan},
  {Ivezi{\'c}}, {Brinkmann}, {Budav{\'a}ri}, {Csabai}, \&
  {Nichol}}]{Richards03}
---. 2003, \aj, 126, 1131

\bibitem[{{Richards} {et~al.}(2006){Richards}, {Strauss}, {Fan}, {Hall},
  {Jester}, {Schneider}, {Vanden Berk}, {Stoughton}, {Anderson}, {Brunner},
  {Gray}, {Gunn}, {Ivezi{\'c}}, {Kirkland}, {Knapp}, {Loveday}, {Meiksin},
  {Pope}, {Szalay}, {Thakar}, {Yanny}, {York}, {Barentine}, {Brewington},
  {Brinkmann}, {Fukugita}, {Harvanek}, {Kent}, {Kleinman}, {Krzesi{\'n}ski},
  {Long}, {Lupton}, {Nash}, {Neilsen}, {Nitta}, {Schlegel}, \&
  {Snedden}}]{Richards06}
---. 2006, \aj, 131, 2766

\bibitem[{{Sandage} {et~al.}(1979){Sandage}, {Tammann}, \& {Yahil}}]{Sandage79}
{Sandage}, A., {Tammann}, G.~A., \& {Yahil}, A. 1979, \apj, 232, 352

\bibitem[{{Schlegel} {et~al.}(1998){Schlegel}, {Finkbeiner}, \&
  {Davis}}]{SFD98}
{Schlegel}, D.~J., {Finkbeiner}, D.~P., \& {Davis}, M. 1998, \apj, 500, 525

\bibitem[{{Schmidt}(1968)}]{Schmidt68}
{Schmidt}, M. 1968, \apj, 151, 393

\bibitem[{{Shankar} \& {Mathur}(2007)}]{Shankar07}
{Shankar}, F., \& {Mathur}, S. 2007, \apj, 660, 1051

\bibitem[{{Silk} \& {Rees}(1998)}]{Silk98}
{Silk}, J., \& {Rees}, M.~J. 1998, \aap, 331, L1

\bibitem[{{Stern} {et~al.}(2005){Stern}, {Eisenhardt}, {Gorjian}, {Kochanek},
  {Caldwell}, {Eisenstein}, {Brodwin}, {Brown}, {Cool}, {Dey}, {Green},
  {Jannuzi}, {Murray}, {Pahre}, \& {Willner}}]{Stern05}
{Stern}, D. {et~al.} 2005, \apj, 631, 163

\bibitem[{{Telfer} {et~al.}(2002){Telfer}, {Zheng}, {Kriss}, \&
  {Davidsen}}]{Telfer02}
{Telfer}, R.~C., {Zheng}, W., {Kriss}, G.~A., \& {Davidsen}, A.~F. 2002, \apj,
  565, 773

\bibitem[{{Vanden Berk} {et~al.}(2001){Vanden Berk}, {Richards}, {Bauer},
  {Strauss}, {Schneider}, {Heckman}, {York}, {Hall}, {Fan}, {Knapp},
  {Anderson}, {Annis}, {Bahcall}, {Bernardi}, {Briggs}, {Brinkmann}, {Brunner},
  {Burles}, {Carey}, {Castander}, {Connolly}, {Crocker}, {Csabai}, {Doi},
  {Finkbeiner}, {Friedman}, {Frieman}, {Fukugita}, {Gunn}, {Hennessy},
  {Ivezi{\' c}}, {Kent}, {Kunszt}, {Lamb}, {Leger}, {Long}, {Loveday},
  {Lupton}, {Meiksin}, {Merelli}, {Munn}, {Newberg}, {Newcomb}, {Nichol},
  {Owen}, {Pier}, {Pope}, {Rockosi}, {Schlegel}, {Siegmund}, {Smee}, {Snir},
  {Stoughton}, {Stubbs}, {SubbaRao}, {Szalay}, {Szokoly}, {Tremonti}, {Uomoto},
  {Waddell}, {Yanny}, \& {Zheng}}]{VandenBerk01}
{Vanden Berk}, D.~E. {et~al.} 2001, \aj, 122, 549

\bibitem[{{Warren} {et~al.}(1994){Warren}, {Hewett}, \& {Osmer}}]{Warren94}
{Warren}, S.~J., {Hewett}, P.~C., \& {Osmer}, P.~S. 1994, \apj, 421, 412

\bibitem[{{Wilkes}(1986)}]{Wilkes86}
{Wilkes}, B.~J. 1986, \mnras, 218, 331

\bibitem[{{Wittman} {et~al.}(2002){Wittman}, {Margoniner}, {Tyson}, {Cohen},
  {Becker}, \& {Dell'Antonio}}]{Wittman02}
{Wittman}, D.~E., {Margoniner}, V., {Tyson}, J.~A., {Cohen}, J.~G., {Becker},
  A., \& {Dell'Antonio}, I.~P. 2002, in Survey and Other Telescope Technologies
  and Discoveries., ed. J.~A. {Tyson} \& .~{Wolff}, S. Proc.~SPIE, 21

\bibitem[{{Wolf} {et~al.}(2003){Wolf}, {Wisotzki}, {Borch}, {Dye},
  {Kleinheinrich}, \& {Meisenheimer}}]{Wolf03}
{Wolf}, C., {Wisotzki}, L., {Borch}, A., {Dye}, S., {Kleinheinrich}, M., \&
  {Meisenheimer}, K. 2003, \aap, 408, 499

\bibitem[{{Wyithe} \& {Loeb}(2003{\natexlab{a}})}]{Wyithe03a}
{Wyithe}, J.~S.~B., \& {Loeb}, A. 2003{\natexlab{a}}, \apj, 586, 693

\bibitem[{{Wyithe} \& {Loeb}(2003{\natexlab{b}})}]{Wyithe03b}
---. 2003{\natexlab{b}}, \apj, 595, 614

\end{thebibliography}
\end{document}